\documentclass[10pt]{article}
\usepackage{amsmath}
\usepackage{amssymb}
\usepackage{amsthm,amsxtra}
\usepackage{epsf}
\usepackage{eepic}
\usepackage{graphicx}
\newlength{\mytopmargin}
\newlength{\myleftmargin}
\usepackage[square,sort,comma,numbers]{natbib}
\numberwithin{equation}{section}

\begin{document}

\title{\Large\bf  Analogies between random
matrix ensembles and the one-component plasma in two-dimensions}
\author{Peter J. Forrester}
\date{}
\maketitle

\begin{center}
\it
Department of Mathematics and Statistics, ARC Centre of Excellence for Mathematical
 and Statistical Frontiers, The University of Melbourne, \\
Victoria 3010, Australia
\end{center}

\bigskip
\begin{center}
\bf Abstract 
\end{center}
\par
\bigskip
\noindent 
{\small The eigenvalue PDF for some well known classes of non-Hermitian random matrices --- the complex
Ginibre ensemble for example --- can be interpreted as the Boltzmann factor for one-component plasma
systems in two-dimensional domains. We address this theme in a systematic fashion, identifying the plasma system
for the Ginibre ensemble of non-Hermitian Gaussian random matrices $G$, the spherical ensemble of the product of
an inverse Ginibre matrix and a Ginibre matrix $G_1^{-1} G_2$, and the ensemble formed by truncating unitary matrices,
as well as for products of such matrices.
We do this when each has either real, complex or real quaternion elements. One consequence of this analogy is that the
leading form of the eigenvalue density follows as a corollary. Another is that the eigenvalue correlations must obey sum
rules known to characterise the plasma system, and this leads us to a exhibit an integral identity satisfied by the
two-particle correlation for real quaternion matrices in the neighbourhood of the real axis.
 Further random matrix ensembles investigated from this viewpoint
are self dual non-Hermitian matrices, in which a previous study has related to the one-component plasma
system in a disk at inverse temperature $\beta = 4$, and the ensemble formed by the single row and column of
quaternion elements from a member of the circular symplectic ensemble.
}

\section{Introduction}

In random matrix theory there are a number of distinguished ensembles --- the classical cases --- for which the
eigenvalue probability density function (PDF) can be calculated explicitly. For example, the classical Gaussian orthogonal ensemble
consisting of real symmetric matrices ${1 \over 2} (X + X^T)$, where $X$ is an $N \times N$ standard real Gaussian matrix,
has its joint eigenvalue PDF proportional to
\begin{equation}\label{1}
\prod_{l=1}^N e^{- {1 \over 2} \lambda_l^2} \prod_{1 \le j < k \le N} | \lambda_k - \lambda_j |.
\end{equation}
This explicit expression was known to Wigner (see \cite{Po65} and references therein).

Wigner \cite{Wi57a} (reprinted in  \cite{Po65})
also observed that (\ref{1}), upon being written in the form $e^{-\beta U}$, is identical to the 
Boltzmann factor for the classical gas in one-dimension with total potential energy
\begin{equation}\label{2}
U = {1 \over 2} \sum_{l=1}^N \lambda_l^2 - \sum_{1 \le j < k \le N} \log | \lambda_k - \lambda_j |, \qquad \lambda_l \in \mathbb R,
\end{equation}
interacting at inverse temperature $\beta = 1$. The first term in (\ref{2}) represents an harmonic attraction towards
the origin, and the second is a pairwise logarithmic repulsion between the particles in the gas. The pair potential 
\begin{equation}\label{Lz}
- \log | z - w|,
\end{equation}
with $z,w \in \mathbb C$, is well known as the solution of the two-dimensional Poisson equation with free boundary
conditions, and thus the pair interaction in (\ref{2}) is that experienced by $N$ two-dimensional unit charges confined to a line.
To understand the origin of the one-body term ${1 \over 2} \lambda^2$ from this perspective, suppose there is a
smeared out neutralising background charge density $-\sigma(\lambda)$, supported on the interval $[-L,L]$. The one-body
term must be the electrostatic energy created by this  background charge, and thus we must have
\begin{equation}\label{2.1}
{1 \over 2} \lambda^2 + C = \int_{-L}^L \sigma(y) \log | \lambda - y| \, dy, \qquad \lambda \in [-L,L],
\end{equation}
for some constant $C$ independent of $\lambda$. The solution of this integral equation, with the requirement that
$\sigma(y)$ vanishes at $y = \pm L$ is (see e.g.~\cite[Prop.~1.4.3]{Fo10})
\begin{equation}\label{2.2}
\sigma(y) = {L \over \pi} \sqrt{ 1 - (y/L)^2}.
\end{equation}
Charge neutrality requires $\int_{-L}^L \sigma(y) \, dy = N$, which in turn implies
\begin{equation}\label{2.2a}
L = \sqrt{2N}.
\end{equation}
In random matrix theory we recognise (\ref{2.2}) and (\ref{2.2a}) as the Wigner semi-circle law for the eigenvalue
density of large random real symmetric matrices; see e.g.~\cite{PS11}. This in fact is one of the derivations of the
law given by Wigner himself \cite{Wi57a}.

Our interest in this paper is in analogies between eigenvalue PDFs with two-dimensional support in the complex plane,
and the Boltzmann factor for one-component log-potential classical gases (plasmas) in two-dimensional domains.
The meaning of the term one-component is that all charges are of the same sign and same strength, $+1$ say.
The specific form of the logarithmic potential will depend on the surface defining the two-dimensional domain,
e.g.~a plane, sphere etc., and the particular boundary condition, which will be either free
or Neumann. As already mentioned,
(\ref{Lz}) applies to planar geometry with free boundary conditions. Since the analogy holds at the level of the PDF,
the corresponding correlations conserve this analogy, and so inherit properties characteristic of the plasma correlations.

The best known example of applicability of the analogy on the random matrix side is the ensemble of $N \times N$ standard
complex Gaussian matrices, also referred to as complex Ginibre matrices \cite{Gi65}. Indeed it was Ginibre who
first computed the joint eigenvalue PDF for this ensemble, showing it to be proportional to
\begin{equation}\label{Z1}
\prod_{l=1}^N e^{- |z_l|^2} \prod_{1 \le j < k \le N} |z_k - z_j|^2, \qquad z_l \in \mathbb C.
\end{equation}
Writing this in the Boltzmann factor form $e^{-\beta U}$ shows
\begin{equation}\label{U1}
U = {1 \over 2} \sum_{l=1}^N |z_l|^2 - \sum_{1 \le j < k \le N} \log | z_k - z_j|
\end{equation}
with $\beta = 2$.
The classical plasma system corresponding to (\ref{U1}) thus consists of $N$ unit charges interacting via the pair potential (\ref{Lz}),
and a one-body two-dimensional harmonic potential towards the origin. The inverse temperature is $\beta =2$.
The one-body potential corresponds to the electrostatic potential
created by a disk of uniform charge density $-1/\pi$ centred on the origin and of radius $\sqrt{N}$. To experience this potential
the particles must be inside the disk; outside the disk the charge density is seen as a point charge of strength $-N$ at the origin,
and the potential  therefore becomes logarithmic. The analogy then breaks down,
 although according to the well known circle law (see e.g.~\cite{TV09}), to leading order all eigenvalues are in the disk of radius $\sqrt{N}$ with
uniform density $1/\pi$. Conversely, as a point to be emphasised in Section \ref{S2}, the fact that a plasma system
will to leading order be charge neutral (if it wasn't, there would be a nonzero electric field and the system
would not be in equilibrium) predicts the circle law for the complex Ginibre matrices.

In Section \ref{S2} we review and extend other examples of the analogy and consequences for the
global eigenvalue distribution. This involves three distinct geometries for the plasma --- the plane, the sphere and
the pseudosphere --- and for each of these geometries three distinct random matrix ensembles depending on the
entries being real, complex or real quaternion.  Products of matrices from these ensembles are also considered from
this viewpoint. In Section \ref{S3} we discuss consequences of the
analogies with respect to general properties of the correlations known as sum rules. Sum rules considered
relate to the moments of the truncated two-particle correlation, the decay of the correlations at the spectrum edge,
and the vanishing of the complex moments of the screening cloud in the cases of fast decay of the two-particle
correlation. The consideration of the latter leads us to exhibit an integral identity satisfied by the limiting 
two-particle correlation for real quaternion Ginibre matrices.
Section \ref{S4} is devoted to providing further evidence to the proposal,
due to Hastings \cite{Ha01}, that with
$$
\mathbb Z_{2N} := \mathbb I_N \otimes  \begin{bmatrix} 0 & - 1 \\ 1 & 0 \end{bmatrix}
$$
and $A_{2N}$ a complex anti-symmetric matrix with each independent element a standard complex Gaussian, the
eigenvalues of the ensemble of random matrices of the form $\mathbb Z_{2N} A_{2N}$ are well described in the large distance
regime by a joint PDF proportional to the form
\begin{equation}\label{Z1x}
\prod_{l=1}^N e^{- |z_l|^2} \prod_{1 \le j < k \le N} |z_k - z_j|^4, \qquad z_l \in \mathbb C.
\end{equation}
In the plasma analogy, the total potential is again given by (\ref{U1}) but with the pre-factor of the one-body term
${1 \over 4}$ instead of ${1 \over 2}$, while the inverse temperature is now $\beta = 4$.
In Section \ref{S5}, the eigenvalue PDF for single (quaternion) row and column deletion of circular symplectic ensemble matrices,
recently computed explicitly in \cite{KK15}, is shown to have a plasma analogue. The latter is known in turn to be
an example of a Pfaffian point process, and this enables the correlation functions for the eigenvalues to be computed,
which in the $N \to \infty$ limit are identified as the eigenvalue correlations for random polynomials with standard complex Gaussian coefficients.

\section{2d plasma analogies for some random matrix ensembles  and consequences for the spectral density}\label{S2}
Characteristic of the eigenvalue PDF for random matrix ensembles is the product of differences as seen in (\ref{1}) and (\ref{Z1}).
Writing the PDF as the exponential of a potential  energy, it is immediate that the corresponding pair potential represents a
logarithmic repulsion between the eigenvalues, as seen from the viewpoint of a classical gas in equilibrium.
A more sophisticated point, already known from the separate works \cite{Kr06}, \cite{FK09}, is that the eigenvalue PDF for particular random matrix ensembles with support in the complex plane naturally projects to certain homogeneous manifolds --- the sphere and the pseudosphere ---
where the eigenvalue density is uniform. Here we collect all these results together, give some extensions,
and we make explicit the plasma prediction
for the eigenvalue density. We remark that the theory of random polynomials, in which the coefficients are complex Gaussians
with certain variances, also permits examples which have uniform density in the plane, on the sphere and on the pseudosphere
\cite{Ha96, EK95,Le00}. 

The
non-Hermitian random matrix ensembles in \cite{Kr06}, \cite{FK09} have complex entries. In these cases,
as well as for the complex Ginibre ensemble, there is also an analogy with the absolute value squared of the wave function
for $N$ spinless fermions subject to a strong perpendicular magnetic field and in the lowest Landau level. This is discussed in
e.g.~\cite[\S 15.2, \S15.6, Ex.~15.7 q.2]{Fo10}. If we consider non-Hermitian random matrix ensemble with real or real quaternion
elements the quantum mechanical analogy breaks down, whereas the plasma analogy persists, now with image charges due to
Neumann boundary conditions. Again this point is known in scattered places in the literature, e.g.~\cite[\S 15.9.1]{Fo10} and
\cite{FM11} in the particular case of the Ginibre ensemble, but has not until now been the subject of a systematic investigation.

\subsection{Ginibre ensembles}
In keeping with Dyson's three fold way \cite{Dy62c}, there are three classes of Ginibre matrices,
i.e.~$N \times N$ Gaussian matrices with all elements independent and thus non-Hermitian. These classes are
distinguished by the number field to which the elements belong --- real, complex or real quaternion. In the latter case
a $2 \times 2$ complex valued matrix representation is used; see e.g.~\cite[\S 1.3.2]{Fo10} and the eigenvalues
occur in complex conjugate pairs. It has already been remarked that the joint eigenvalue PDF in the complex case is given by (\ref{1}), and the analogy
with a one-component plasma with neutralising background in a uniform disk of radius $\sqrt{N}$ has been noted.
Here we want to emphasise the reasoning which enables the leading order particle density of the plasma system,
and thus the leading order eigenvalue density of the random matrix ensemble, to be determined.

The first step is to introduce a so-called global scaling of (\ref{Z1}) by replacing $z_l \mapsto \sqrt{N} z_l$.
A crucial feature of the product of differences in (\ref{Z1}) is its scale invariance, changing by a multiplicative
factor only. Introducing too the empirical density $\hat{\sigma}(z) :=
\sum_{p=1}^N \delta ( z_p - z)$ shows that (\ref{Z1}) is then proportional to
\begin{equation}\label{2.t}
\exp \Big (2\Big  \{  - N \int_{\mathbb C} \hat{\sigma}(z)  | z|^2 \, d^2 z + {1 \over 2}
\int_{\mathbb C \times \mathbb C \backslash \{ z = w\}} d^2 z \, d^2 w \,
 \hat{\sigma}(z)  \hat{\sigma}(w)  \log | z - w|  \Big \} \Big ).
 \end{equation}
We remark that the excluded set $z=w$ in the double integral in (\ref{2.t}) corresponds, in physical terms, to the
self energy of the charge density $\hat{\sigma}(z)$.

The essential physical idea, made rigorous using the theory of large deviations \cite{PH98,bAZ97} (see the discussion in the introduction
to \cite{SS15}) is that for large $N$ the empirical density can be replaced by $N \rho^{\rm g}(z)$, where $\rho^{\rm g}(z)$ is the
limiting global particle density normalised to have total integral equal to unity. Doing this, (\ref{2.t}) reads
\begin{equation}\label{2.ta}
\exp \Big ( - 2 N^2 \mathcal E [\rho(z)] \Big ), \qquad 
 \mathcal E [\rho(z)] := -  {1 \over 2} \int_{\mathbb C} {\rho}^{\rm g}(z)  | z|^2 \, d^2 z + {1 \over 2}
\int_{\mathbb C \times \mathbb C} d^2 z \, d^2 w \,
 \rho^{\rm g} (z)  \rho^{\rm g}(w)  \log | z - w|  .
 \end{equation} 
 Note that in this double integral there is no need to exclude the self energy term, as in contrast to (\ref{2.t}) where it diverges,
 it makes a vanishingly small contribution. The final ingredient, which in idea is classical going back to at least
 Gauss (see e.g.~ \cite{ST98}), is that the global density in (\ref{2.ta}) is such that the energy functional 
  $\mathcal E [\rho(z)]$ is minimised. From the calculus of variations, this occurs when
\begin{equation}\label{2.tb}  
0 = - {1 \over 2} |z|^2 + \int_{\mathbb C} d^2 w \,  \rho^{\rm g}(w)  \log | z - w| .
\end{equation}
But fundamentally $-\log|z - w|$ is the solution of the two-dimensional Poisson equation
$\nabla^2_z \phi(z,w) = - 2 \pi \delta^{(2)}(z - w)$ in free boundary condition, so acting on
(\ref{2.tb}) with $\nabla^2$ tells us that $\nabla^2 |z|^2 = 2 \pi  \rho^{\rm g}(z)$, 
for $z \in \mathcal D$. The region $\mathcal D$ is referred to as the droplet
(see e.g.~\cite{ZW06}), and hence
inside this region we have $ \rho^{\rm g}(z) = {1 \over \pi}$, while $ \rho^{\rm g}(z)$ vanishes
outside $\mathcal D$. The last point to note is that to minimise $\mathcal E [\rho(z)]$, inspection of the
functional form requires $\mathcal D$ to be a disk about the origin, and the requirement that
the integral of  $\rho^{\rm g}(z)$ over $\mathcal D$ equals unity tells us that in fact
$\mathcal D$ is the unit disk. The unit disk is precisely the scaled limit 
of the uniform background
charge density which gave rise to the Boltzmann factor (\ref{Z1}) in the plasma interpretation. 

Let's consider next the case of real quaternion elements. The eigenvalue PDF for the eigenvalues in the upper half
complex plane is then proportional to \cite{Gi65}, \cite{Me91}
\begin{equation}\label{N1}
\prod_{j=1}^N e^{- 2 |z_j|^2} |z_j - \bar{z}_j|^2 \prod_{1 \le j < k \le N} |z_k - z_j|^2 | z_k - \bar{z}_j|^2.
\end{equation}
A plasma analogy has previously been identified in \cite[Prop.~15.9.1]{Fo10}. To review this result, the first point to
note is that the solution of the two-dimensional Poisson equation $\nabla^2_{\vec{r}} \phi(\vec{r}, \vec{r}\,') = - 2\pi \delta(\vec{r} - \vec{r}\,')$
with the Neumann boundary condition along the $x$-axis
\begin{equation}
{\partial \over \partial y}  \phi(\vec{r}, \vec{r}\,') \Big |_{y \to 0^+} = 0,
\end{equation}
is, using complex coordinates, given by
\begin{equation}\label{pp}
 \phi(\vec{r}, \vec{r}\,') = - \log \Big ( | z - z'| \, |z -\bar{z}\,'| \Big ).
 \end{equation}
As discussed in \cite[\S 15.9]{Fo10}, this formally occurs when the region $y < 0$ has zero dielectic constant, which effectively gives
rise to an image particle of identical charge at the reflection point $\bar{z}\,'$ of $z\,'$ about the real axis. We say formally, since physically a
dielectric constant cannot be less than its vacuum value of unity. On the other hand, this circumstance is mimicked when the dielectric constant of the
region $y > 0$ is much greater than the dielectric constant for $y < 0$.

Following \cite[Prop.~15.9.1]{Fo10}, consider now a one-component system --- all particles of unit charge --- confined to the semi-disk $|z| < \sqrt{N}$, $y > 0$, with a neutralising
background of uniform density $\rho = 2/\pi$ filling the semi-disk. With the system interacting at the inverse temperature $\beta$, a short computation shows that the corresponding Boltzmann
factor is proportional to
\begin{equation}\label{N2}
\prod_{j=1}^N e^{-\beta |z_j|^2} |z_j - \bar{z}_j|^{\beta / 2} \prod_{1 \le j < k \le N} |z_k - z_j|^\beta
|z_k - \bar{z}_j|^\beta.
\end{equation}
In this expression, the origin of the term $\prod_{j=1}^N |z_j - \bar{z}_j|^{\beta / 2}$ requires further explanation. The required theory is that the
self-energy term in the total potential is computed according to the formula
\begin{equation}\label{N2s}
{1 \over 2} \sum_{j=1}^N \lim_{z \, ' \to z_j} \Big (
\phi(z_j,z') - \log |z_j - z| \Big ) = - {1 \over 2} \sum_{j=1}^N  |z_j -\bar{z}_j|.
\end{equation}
Note in particular the factor of ${1 \over 2}$, which can be interpreted as saying the self energy is equally distributed between
the actual and image system.

Comparing the eigenvalue PDF (\ref{N1}) and the Boltzmann factor 
(\ref{N2}) shows that for $\beta = 2$ they are identical, provided that in the Boltzmann factor an
additional self-energy factor is included. Such a self-energy term is not expected to effect the large distance behaviours of the plasma,
which instead are determined by the pair potential. 
Specific consequences by way of sum rules will be discussed in the next section. In this section the consequence of the plasma analogy that
we emphasise is the implied eigenvalue density. 
By construction of the Boltzmann factor, the background charge density minimises the energy functional.
Thus 
to leading order the eigenvalue density will be given by the background density,
and so will be uniform in the semi-disk $|z| < \sqrt{N}$, $y > 0$.

The case of real Ginibre matrices is more complicated than for the complex or real quaternion Ginibre matrices.
This is because in the real case there is a non-zero probability of some eigenvalues being real. Consequently the
joint eigenvalue PDF consists of disjoint sectors depending of the number of real eigenvalues, $k$ say
($k$ must be of the same parity as $N$). In the upper half plane --- note that all entries being real implies all the
complex eigenvalues occur in complex conjugate pairs --- the joint eigenvalue PDF, conditioned to have
precisely $k$ real eigenvalues $\{\lambda_l\}_{l=1,\dots,k} $, is proportional to \cite{Ed97}
\begin{eqnarray}\label{3.1}
&& {1 \over 2^{N(N+1)/4} \prod_{l=1}^N \Gamma(l/2) }
{2^{(N-k)/2} \over k! ((N-k)/2)! } \Big | \Delta(\{\lambda_l\}_{l=1,\dots,k} \cup
\{ x_j \pm i y_j \}_{j=1,\dots,(N-k)/2}) \Big | \nonumber \\
&& \qquad \times
e^{- \sum_{j=1}^k \lambda_j^2/2} e^{\sum_{j=1}^{(N-k)/2}(y_j^2 - x_j^2)}
\prod_{j=1}^{(N-k)/2} {\rm erfc}(\sqrt{2} y_j),
\end{eqnarray}
where $\Delta(\{z_p\}_{p=1,\dots,m}) := \prod_{j < l}^m (z_l - z_j)$. Here 
$\lambda_l \in (-\infty, \infty)$ while $(x_j,y_j) \in \mathbb R \times {\mathbb R}_+$,
${\mathbb R}^2_+ := \{ (x,y) \in {\mathbb R}^2 : \, y>0 \}$. 

We can write
\begin{multline}\label{SAS}
 \Big | \Delta(\{\lambda_l\}_{l=1,\dots,k} \cup
\{ x_j \pm i y_j \}_{j=1,\dots,(N-k)/2}) \Big | \\
 =
\exp \Big ( - \sum_{1 \le j < p \le k} \log | \lambda_p - \lambda_j | -
\sum_{j=1}^k \sum_{s=k+1}^{(N+k)/2} \log |z_s - \lambda_j|   | \bar{z}_s - \lambda_j| -
\sum_{a,b = k+1}^{(N+k)/2} \log |z_a - \bar{z}_b| \Big ) \\
\times
\exp \Big ( - \sum_{k+1 \le a < b \le (N+k)/2} \log | z_b - z_a|  | \bar{z}_b - \bar{z}_a | \Big ).
\end{multline}
A plasma interpretation of (\ref{SAS}) has been discussed previously \cite{FM11} in an analogous case (real spherical ensemble; see the next subsection).
There the interpretation given was entirely in terms of the pair potential (\ref{Lz}), with image charges imposed as a constraint.
An alternative viewpoint is to consider instead the pair potential (\ref{pp}), with the charges on the real axis of strength $1/2$, at inverse
temperature $\beta = 2$.
Turning now to the one body term in (\ref{3.1}), we first make the manipulation
\begin{equation}\label{see}
e^{- \sum_{j=1}^k \lambda_j^2/2} e^{\sum_{j=1}^{(N-k)/2}(y_j^2 - x_j^2)}
\prod_{j=1}^{(N-k)/2} {\rm erfc}(\sqrt{2} y_j) =
e^{- \sum_{j=1}^k \lambda_j^2/2} e^{- \sum_{j=1}^{(N-k)/2}(x_j^2 + y_j^2)}
\prod_{j=1}^{(N-k)/2} e^{2 y_j^2} {\rm erfc}(\sqrt{2} y_j).
\end{equation}
We see that the first two exponential terms on the RHS result from a coupling between the charges 
confined to the semi-disk $|z| < \sqrt{N}$, $y > 0$, 
and with a neutralising
background of uniform density $\rho = 1/\pi$ filling the semi-disk (this is half the value of the corresponding background
density in the real quaternion case due to there being of order $N/2$ rather than of order $N$ eigenvalues in the semi-disk).
The final term can be interpreted as the coupling to some external short range potential --- note that to leading order
it tends to unity as $y_j \to \infty$.

Analogous to the circumstance already mentioned in the case of the complex Ginibre ensemble, the neutralising background being the semi-disk 
$|z| < \sqrt{N}$, $y > 0$, implies that to leading order the particle density will also be uniform in this region. For the real Ginibre ensemble,
the eigenvalues not on the real axis occur in complex conjugate pairs, so the corresponding eigenvalue density is to leading order
anticipated to be uniform in the disk $|z| < \sqrt{N}$ in keeping with the circle law. Note in particular that the real eigenvalues play
no role in this leading order statement, as their proportion is on average $1/\sqrt{N}$ \cite{Ed97}, and is thus vanishingly small as
$N \to \infty$.

Let $G$ be from a Ginibre ensemble (real, complex or real quaternion entries), and construct the matrix \cite{SASS88}
\begin{equation}\label{YG}
Y = {1 + \sqrt{c} \over 2} G + {1 - \sqrt{c} \over 2} G^\dagger, \qquad c = {1 - \tau \over 1 + \tau} \: \: (0 \le \tau < 1).
\end{equation}
The eigenvalue PDF of this deformation of the Ginibre ensemble was analysed in \cite{LS91} in the real case, in \cite{FKS97} in the complex
case, and in \cite{Ka98} in the real quaternion case. In each case, the only modification to the eigenvalue PDF relative to that
for the case $c = 1$, i.e.~the original Ginibre ensemble, occurs in the one-body factors. Specifically, in the
complex and real quaternion cases, replace
\begin{equation}\label{p.s}
e^{- \sum_{j=1}^N | z_j|^2} \mapsto \exp \Big ( - {1 \over 1 - \tau^2} \sum_{j=1}^N \Big ( |z_j|^2 - {\tau \over 2} (z_j^2 +
\bar{z}_j^2 ) \Big ) \Big ),
\end{equation}
while in the real case (\ref{SAS}), replace
$$
\prod_{j=1}^{(N-k)/2} {\rm erfc}(\sqrt{2} y_j) \mapsto \prod_{j=1}^{(N-k)/2}  {\rm erfc} \Big ( {\sqrt{2} \over 1 - \tau} y_j \Big ),
$$
up to the normalisation. In the real case, substituting in (\ref{SAS}) and replacing the complimentary error function
by its leading asymptotic form reclaims the RHS of (\ref{p.s}) for the functional form of the one-body
factor for the complex eigenvalues.

Thus, from the viewpoint of the present theme, our task is to interpret the RHS of (\ref{p.s}) as a coupling
of a background charge density to the mobile charges of the plasma. Actually, the answer to this is already
known \cite{DGIL94,FJ96}; see also \cite[Ex.~15.2 q.4]{Fo10}. We have that the  RHS of (\ref{p.s})  results from a uniformly
charged ellipse, charge density $-1/\pi (1 - \tau^2)$, with semi-axes $A = \sqrt{N}(1 + \tau)$,
$B = \sqrt{N}(1 - \tau)$. Note that this is consistent with the elliptic law in random matrix theory
(see e.g.~\cite{NO14}) as a generalisation of the circular law.

\subsection{Spherical ensembles}
Following \cite{Kr06}, as motivation for the study of the spherical ensembles as a natural next step from the study of the
Ginibre ensembles, take two Ginibre matrices $G_1$ and $G_2$, both with entries from the same number field
and consider the generalised eigenvalue problem $G_1 \vec{v} = \lambda G_2 \vec{v}$. The generalised eigenvalues $\lambda$
are the eigenvalues of the random matrix product $G_2^{-1} G_1$, and it is eigenvalues of this type of matrix that make up the spherical
ensemble.

The naming can be justified by relating the eigenvalue distribution of  $G_2^{-1} G_1$ to the geometry of the sphere. Further following \cite{Kr06},
for this purpose we specialise to the complex case, and introduce the transformed pair of matrices $(C,D)$ according to
$$
C := -\bar{\beta} G_2 + \bar{\alpha} G_1, \qquad D = \alpha G_2 + \beta G_1,
$$
with $\alpha, \beta \in \mathbb C$ and such that 
\begin{equation}\label{ab}
|\alpha|^2 + |\beta|^2 = 1. 
\end{equation}
One can check that $(G_1,G_2)$ has the same
distribution as $(C,D)$. From this latter fact one can check that the distribution of the generalised eigenvalues  is
unchanged by the fractional linear transformation
\begin{equation}\label{ab1}
\lambda \mapsto {\lambda \alpha - \bar{\beta} \over \lambda \beta - \bar{\alpha}}.
\end{equation}
This combined with (\ref{ab}) implies that upon a stereographic projection from the complex plane to the
Riemann sphere the eigenvalue distribution of $G_2^{-1} G_1$ is invariant under rotation of the sphere, and
thus the name of the ensemble.

The plasma analogy is very direct in the complex case. Thus in this case the joint eigenvalue PDF  is
proportional to \cite{Kr06}
\begin{equation}\label{CV}
\prod_{l=1}^N {1 \over (1 + |z_l|^2)^{N+1}} \prod_{1 \le j < k \le N} |z_k - z_j|^2, \qquad z_l \in \mathbb C.
\end{equation}
As noted in e.g.~\cite[\S 15.6.4]{Fo10}, a stereographic projection from the Riemann sphere of radius ${1 \over 2}$ to the complex
plane, located  tangent to the north pole, is specified by the equation
\begin{equation}\label{SP}
z = e^{i \phi} \tan {\theta \over 2}.
\end{equation}
Making this change of variables transforms (\ref{CV}) to the PDF on the sphere proportional to
\begin{equation}\label{CV1}
\prod_{1 \le j < k \le N} |\vec{r}_k - \vec{r}_j|^2,
\end{equation}
where $\vec{r}_j$ is the vector in $\mathbb R^3$ corresponding to the point $(\theta_j,\phi_j)$ on the sphere.

To compare (\ref{CV1}) to the Boltzmann factor for a one-component plasma on a sphere, the first point to note is that 
in this geometry the solution of the charge neutral Poisson equation
\begin{equation}\label{SSP}
\nabla^2_{\theta, \phi} \phi((\theta, \phi), (\theta',\phi')) = - 2 \pi \delta_S((\theta,\phi), (\theta',\phi')) + {1 \over 2 R^2}
\end{equation}
(charge neutrality is a necessary condition for the existence of a solution, due to the sphere being a compact surface),
where $\delta_S((\theta, \phi), (\theta',\phi'))$ is the Dirac delta function on the sphere and $R$ is the radius, is given by
\begin{equation}\label{so}
- \log |\vec{r} - \vec{r}\,'|, 
\end{equation}
 where $\vec{r}, \vec{r}\,'$ are the points in $\mathbb R^3$ corresponding to
$(\theta, \phi)$, $ (\theta',\phi')$ on the sphere. Hence for $N$ unit charges  on the sphere,
in the presence of a uniform background, and
at inverse temperature $\beta = 2$, the Boltzmann factor is precisely (\ref{CV1}) \cite{Ca81}. The uniform background
of the plasma is consistent with the uniform eigenvalue density when projected on the sphere,
as follows from the invariance (\ref{ab1}).

In the case of the real quaternion spherical ensemble, the joint eigenvalue PDF in the complex plane is
proportional to \cite{Ma13}
\begin{equation}\label{2.12a}
\prod_{j=1}^N {|\lambda_j - \bar{\lambda}_j|^2 \over (1 + |\lambda_j|^2)^{2(N+1)}}
\prod_{1 \le j < k \le N} |\lambda_k - \lambda_j|^2 | \lambda_k - \bar{\lambda}_j|^2, \qquad \lambda_j \in \mathbb C_+.
\end{equation}
Upon stereographic projection, this maps to the PDF on the half sphere $\mathbb  S_+^2$, the latter defined by the restriction on the azimuthal
angle $0 \le \phi \le \pi$, proportional to
\begin{equation}\label{CV2}
\prod_{j=1}^N | \vec{r}_j - \vec{r}_j^{\, *}|^2
\prod_{1 \le j < k \le N} |\vec{r}_k - \vec{r}_j|^2 |\vec{r}_k - \vec{r}_j^{\,*}|^2, \qquad \vec{r}_j \in \mathbb S_+^2,
\end{equation}
where $\vec{r}_j^{\, *}$ is the vector in $\mathbb R^3$ corresponding to the point on the sphere $(\theta_j, - \phi_j)$.
The plasma interpretation is the sphere analogue of that for the real quaternion Ginibre eigenvalue PDF (\ref{N1}).
Thus one considers $N$ unit charges confined to the half sphere $\mathbb  S_+^2$, with Neumann boundary conditions at the boundary,
for which the potential at $\vec{r}_b$ due to a charge at $\vec{r}_a$ is
\begin{equation}\label{vab}
- \log \Big ( | \vec{r}_b - \vec{r}_a| | \vec{r}_b - \vec{r}_a^{\, *}| \Big ),
\end{equation}
and in the presence of a neutralising background.
At the coupling $\beta = 2$ the corresponding Botzmann factor is (\ref{CV2}), except that in (\ref{CV2}) there is an additional
one-body factor $\prod_{j=1}^N | \vec{r}_j - \vec{r}_j^{\, *}|$. The background being uniform is the plasma mechanism
for the eigenvalue density also being uniform to leading order as $N \to \infty$ \cite{Ma12}.

The real spherical ensemble shares with the real Ginibre ensemble the property of having real eigenvalues, in addition to
eigenvalues occurring in complex conjugate pairs, and the joint eigenvalue PDF correspondingly breaks up into sectors
analogous to (\ref{3.1}).  The functional form of the one-body terms are at first complicated \cite[Eqns.~(21) and (22)]{FM11},
but simplify upon a fractional linear transformation mapping the upper half plane to the unit disk \cite[Eqns.~(6) and (7)]{FM11}.
Then mapping the unit disk to the upper half of the Riemann sphere $0 < \theta < \pi/2$ via a stereographic projection with
the complex plane passing through the sphere at the equator, we see that the effective pair potential for the complex eigenvalues
is again (\ref{vab}), but now with $\vec{r}^{\, *}$ obtained from $\vec{r}$ by mapping the polar angle $\theta \mapsto \pi - \theta$. The real eigenvalues
are their own images, and to account for this it is necessary that they be
chosen to have charge $1/2$ rather than unity as for the complex eigenvalues. Some one-body terms remain, but they
are independent of $N$, and so are not expected to alter and large distance properties of the plasma. In particular there being a uniform
neutralising background tells us that to leading order in $N$ the eigenvalue density will be uniform when projected on the sphere, which is indeed
the case \cite{FM11,Ca81}. As for the real Ginibre ensemble, the fact that there are real eigenvalues does not effect this property.

\subsection{Truncated unitary ensembles}\label{S2.3}
The eigenvalue distribution in the complex plane for sub-matrices of unitary matrices was first considered by
 \.Zyczkowski and  Sommers \cite{ZS99}. The setting is to choose a matrix from the set of $(N+n) \times (N+n)$ unitary matrices
 ${\rm U}(N+n)$ under the assumption
 of Haar measure, then to delete $n$ rows and $n$ columns to form an $N \times N$ sub-matrix. The point of interest is the corresponding
 eigenvalue PDF, which was shown in \cite{ZS99} to be proportional to 
 \begin{equation}\label{D1}
 \prod_{l=1}^N(1 - |z_l|^2)^{n-1} \chi_{|z_l|<1} \prod_{1 \le j < k \le N} | z_k - z_j|^2,
 \end{equation}
where $\chi_A = 1$ for $A$ true and $\chi_A = 0$ otherwise. The plasma analogy was identified in \cite{FK09}, and now relates
not to the sphere but to the pseudosphere, which is a two-dimensional hyperbolic space with constant negative Gaussian
curvature $\kappa = - 1/a^2$. It is naturally embedded in three-dimensional Minkowski space, with co-ordinates
$(y_0,y_1,y_2)$ and line element such that $(ds)^2 = - (dy_0)^2 + (dy_1)^2 + (dy_2)^2$. Specifically, the pseudosphere is
the branch of the solution of the equation $-y_0^2 + y_1^2 + y_2^2 = - a^2$ which includes the point $(a,0,0)$. This branch
can be parameterised by
\begin{equation}\label{bb}
y_0 = a \cos \tau, \qquad y_1 = a \sinh \tau \cos \phi, \qquad y_2 = a \sinh \tau \sin \phi.
\end{equation}
With $z = x+iy$, the pseudosphere is projected onto the Poincar\'e disk via the stereographic projection (as interpreted in Euclidean space)
\begin{equation}\label{bb1}
z = 2a \tanh {\tau \over 2} \, e^{i \phi}, \qquad |z| < 2a,
\end{equation}
(we will take $a=1/2$)
or equivalently by the polar form $z = r e^{i \phi}$ with $r = 2a \tanh {\tau \over 2}$. Note that this is identical to (\ref{SP}), if we set $\tau = i \theta$,
$a = iR$, where $R$ is the radius of the sphere (in (\ref{SP}), $R=1/2$), thus justifying the name pseudosphere. It turns out that this
same prescription applies to deducing the solution of the Poisson equation on the pseudosphere from knowledge of the solution
of (\ref{SSP}), giving the pair potential (see e.g.~\cite{JT98})
 \begin{equation}\label{D2}
 - \log \Big ( {|z_j - z_k| \over (1 - r_j^2)^{1/2}  (1 - r_k^2)^{1/2} } \Big ).
 \end{equation}

A significant difference between the pseudosphere and the sphere is that the former is not compact, and so does not require
charge neutrality on the RHS of the Poisson equation for a solution. This is just as well, since setting $R=ia$ in (\ref{SSP})
implies that the smeared out background is of the same sign as the mobile charges, with charge density $1/4 \pi a^2$. With
$N$ particles, the total charge density is $N/4 \pi a^2$, so to get a net background charge density $-\eta$ we must cancel
this and impose a smeared out charge density $-\eta - N/ 4 \pi a^2$. A short computation \cite{JT98}  gives that
for $\beta = 2$ and $a=1/2$, and using the variables (\ref{bb1}),
 the corresponding Boltzmann factor is proportional to
\begin{equation}\label{D3}
\prod_{j=1}^N \Big ( 1 - |z_j|^2 \Big )^{\pi \eta + 1} \chi_{|z_j| < 1} \prod_{1 \le j < k \le N} | z_k - z_j|^2,
\end{equation}
and hence is the same functional form as the PDF (\ref{D1}) with 
\begin{equation}\label{D4}
n = \pi \eta + 2.
\end{equation}

We can use the plasma analogy to predict the leading order large $N$ form of the eigenvalue density,
in the case that $n/N \to c > 0$. On the pseudosphere the neutralising background has constant charge
density $-\eta$, and we have $\eta \approx n/\pi$ for $n$ large. We know too (see e.g.~\cite[Eq.~(15.161)]{Fo10})
that the pseudosphere surface element $dS$ projects to the Poincar\'e disk according to
\begin{equation}\label{SS}
dS = {dx dy \over (1 - |z|^2)^2}.
\end{equation}
The total number of particles in a disk of radius $R$ within the Poincar\'e disk is thus $\pi \eta R^2/(1 - R^2)$. Substituting for $\eta$
and equating this to $N$ allows us to specify $R$, and we conclude that the leading order eigenvalue density will be given by
\begin{equation}\label{RR}
{n + N \over \pi (1 - |z|^2)^2} \chi_{|z| < R}, \qquad  R =  \Big ( {1 \over 1 + n/N } \Big )^{1/2},
 \end{equation}
 in agreement with the result of an explicit
analysis of the one-point density \cite{FS03}.

We now turn our attention to the case of truncations of unitary matrices with real quaternion elements.
Such matrices are equivalent to unitary symplectic matrices, and so make up one of the classical groups,
denoted Sp$(2m)$ for matrices of size $m \times m$ (the $2m$ in Sp$(2m)$ refers to the row or column size
of the corresponding complex matrix --- recall each real quaternion is represented as a $2 \times 2$ matrix).
We suppose that the original matrices are of size $(N+n) \times (N+n)$, and consider sub-blocks of size $N \times N$.
The corresponding eigenvalue PDF has been reported in the recent work \cite{IK14}, although no details
as to its derivation were given. As a possible strategy,
we begin by noting from \cite[$N \mapsto N + n, n_1 = n_2 = N, \beta = 4$]{Fo06}
that the distribution of an $N \times N$ sub-block $A$ say is proportional to
\begin{equation}\label{F1}
\det ( \mathbb I_N - A^\dagger A)^{2 (n-N+1/2)}.
\end{equation}
Note that this requires $n \ge N$ to be well defined; if not some eigenvalues of $A^\dagger A$ are unity, and the distribution of $A$
is singular.

Next we note that the distribution (\ref{F1}) is analogous to the distribution
\begin{equation}\label{Aq}
\det ( \mathbb I_N + A^\dagger A)^{- 2 (n + N)}
\end{equation}
for matrices $A = (B^\dagger B)^{-1/2} X$, with $B,X$ given by $n \times n$ ($n \times N$)
real quaternion matrices respectively, for which Mays \cite{Ma12,Ma13} has computed the joint eigenvalue
PDF as proportional  to (\ref{2.12a}) but with the replacement
\begin{equation}\label{2.13b}
1/(1 + |\lambda_j|^2)^{2(N+1)} \mapsto  1/(1 + |\lambda_j|^2)^{2(n+1)}.
\end{equation}
The significance of this is that, as first demonstrated in the complex case, and used
subsequently in the real case \cite{Ma12}, the working required to deduce the joint eigenvalue PDF
starting with a spherical ensemble density such as (\ref{Aq}) is structurally identical to that required for the
same task as starting with the PDF (\ref{F1}). The final results must be the same except that the
parameter $n$ is to be replaced by $-(n+1/2)$ as is consistent in going from (\ref{Aq}) to (\ref{F1}),
and the factors $(1+|\lambda_j|^2)$ must be replaced by $(1 - |\lambda_j|^2)$ for the same reason.
Hence, from knowledge of the eigenvalue PDF corresponding to (\ref{Aq}) being given by (\ref{2.12a}) with the
replacement (\ref{2.13b}), it must be that the eigenvalue PDF corresponding to the distribution 
(\ref{F1}) on random matrices with real quaternion entries is proportional to
\begin{equation}\label{2.13c}
\prod_{l=1}^N (1 - |z_l|^2)^{2n-1} | z_l - \bar{z}_l|^2
\prod_{1 \le j < k \le N} |z_k - z_j|^2 | z_k - \bar{z}_j|^2, \quad |z_l| < 1,
\end{equation}
in agreement with the functional form reported in \cite{IK14}.
As in the complex case \cite{FK09}, even though our derivation of (\ref{2.13c}) has required $n \ge N$ due to it being
based on (\ref{F1}), the final expression is well defined for all $n \ge 1$ and is expected to be generally true (in the complex
case this can be checked by an alternative derivation \cite{ZS99}).

For the plasma analogy, we consider a half pseudosphere, specified by (\ref{bb}) with $0 \le \phi \le \pi$ and impose
Neumann boundary conditions at $\phi = 0, \pi$. The corresponding pair potential is then
\begin{equation}\label{D2a}
 - \log \Big ( {|z_j - z_k| |z_j - \bar{z}_k| \over (1 - r_j^2)  (1 - r_k^2) } \Big )
 \end{equation}
(cf.~(\ref{D2})). Recalling that the solution (\ref{D2a}) corresponds to setting $R=ia$ in (\ref{SSP}), and thus each charge effectively
contributes uniform smeared out background of charge density $1/4 \pi a^2$, we see that we must impose a smeared out
background charge density $-2(\eta + N/4 \pi a^2)$ on the half pseudosphere. As noted in \cite{JT98}, a
one-body potential $V(r)$ satisfying
$$
\nabla^2 V = 4 \pi (\eta + N/4 \pi a^2)
$$
is then created, where $\nabla^2$ is the appropriate Laplacian operator on the pseudosphere. This has solution
$$
V(r) = - (4 \pi \eta a^2 + N) \log \Big ( 1 - {r^2 \over 4 a^2} \Big ),
$$
and the total potential energy due to the coupling between the background and the particles is then $\sum_{j=1}^N V(r_j)$.
Adding to this the potential energy due to the particle-particle interaction as calculated from the pair potential (\ref{D2a}),
and the self energy terms as calculated using the LHS of (\ref{N2s}), we see that the Boltzmann factor is proportional to
\begin{equation}\label{LL1}
\prod_{j=1}^N (1 - |z_j|^2)^{2 \pi \eta + 1}
|z_j - \bar{z}_j| \chi_{|z_j| < 1}
\prod_{1 \le j < k \le N} |z_k - z_j|^2 |z_k - \bar{z}_j |^2.
\end{equation}
As in common with (\ref{N1}) and (\ref{CV2}), the eigenvalue PDF (\ref{2.13c}) contains an extra 
factor of the term in (\ref{LL1}) corresponding to the self-image. Also, the identification (\ref{D4}) must
now be modified to read
\begin{equation}\label{D4a}
n = \pi \eta + 1.
\end{equation}
With these points noted, we thus have an analogy between the eigenvalue PDF for truncations of random unitary
matrices with real quaternion elements, and the Boltzmann factor of the one-component plasma on the half pseudosphere
with Neumann boundary conditions. As in the complex case, an immediate consequence, due to the background being
uniform, is the formula (\ref{RR}) for the predicted leading order eigenvalue density in the case that $n,N \to \infty$ with
$n/N$ fixed.

The final case to consider relating to truncations is when the elements of the unitary matrix are real, and so the
matrices a real orthogonal, which like Sp$(2m)$ make up one the classical groups, the orthogonal group denoted
${\rm O}(m)$ for matrices of size $m \times m$.
It is shown in \cite{KSZ09} that
the joint eigenvalue PDF,
which due to there being real eigenvalues breaks up into sectors depending on the number of real eigenvalues,
is structurally identical to (\ref{3.1}) but with the one body factors
(\ref{see}) replaced by $\prod_{l=1}^k w_r(\lambda_l) \prod_{s=1}^{(N-k)/2} w_c(x_s,y_s)$, where with $z = x + i y$,
\begin{equation}\label{ug}
w_r(\lambda) \propto (1 - \lambda^2)^{n/2 - 1}, \quad
w_c(x,y) \propto |1 - z^2 |^{n-2} \Big ( |1 - z^2 |^{n/2-1}
 \int_{2y/|1 - z^2|}^1 (1 - t^2)^{(n-3)/2} \, dt \Big )^2.
\end{equation}
We have discussed the plasma interpretation  of the two body factors in (\ref{3.1}), which are thus the same in the
present setting. For the one-body terms, we note that the factor in $w_c(x,y)$ in large brackets is of order unity
as $|z| \to 1$ and so can be considered as a coupling to a short range potential, as with the final term in (\ref{see}),
whereas the first factor in $w_c(x,y)$ and the expression for $w_r(\lambda)$ each result from a coupling between
a neutralising background $-(\eta + N/4 \pi a^2)$ on the half pseudosphere, and the pair potential (\ref{D2a})
for a unit and half unit charge respectively. A precise relation between $\eta$ and $n$ as in (\ref{D4}) and (\ref{D4a})
cannot be made due to the arbitrariness of the factorisation of $w_c(x,y)$. But for large $n$ we must have have
$n \approx \pi \eta$ and this implies the validity of the plasma analogy prediction of the spectral density again being
given by (\ref{RR}), in the appropriate limit.

\subsection{Induced ensembles}

Let $G$ be an $N \times N$ random matrix with real, complex or real quaternion  elements, and let $U$ be a unitary matrix
with elements from the same number field. With 
\begin{equation}\label{gA}
A = (G^\dagger G)^{1/2} U, 
\end{equation}
it follows that $A^\dagger A$ 
and $G^\dagger G$ have the same matrix distribution, and 
in particular that $A^\dagger A$ and $G^\dagger G$ have the same distribution of singular values.
Suppose that in addition
the distribution of $G$ is unchanged by left or right multiplication by a unitary matrix. Since by the singular value decomposition,
for some unitary matrices $U_1, U_2$ we have
$G = U_1 \Sigma U_2$, where $\Sigma$ is a diagonal matrix of the singular values, by this last assumption $G$ has the same
distribution as $U \Sigma $, where $U$ is chosen with Haar measure.
Substituting this in the definition of $A$ tells us that $A$ has the same distribution as $\Sigma U$, but this is also distributed as $G$,
so we come to the conclusion that $A$ and $G$ are equal is distribution, and so have the same distribution of eigenvalues.
Note in particular that the Ginibre, spherical and truncated unitary ensembles all have their elements $G$ unchanged in distribution by
left or right multiplication by a unitary matrix, and so this statement applies to those cases.

The construction (\ref{gA}) is well defined for $G$ rectangular of size $n \times N$ say.
Given that $G$ is from a random matrix ensemble, a relevant question to ask is how the volume element of $G$,
$(dG)$, is related to $(dA)$. Following \cite{FBKSZ12}, to answer this question we begin by considering $C := G^\dagger G$.
With
$n \ge N$, it is a standard result (see e.g.~\cite[Eq.~(3.30)]{Fo10}) that
\begin{equation}\label{D1j}
(dG) \propto (\det C)^{(\beta/2)(n-N-1+2/\beta)}(dC),
\end{equation}
where $\beta = 1,2,4$ according to the entries of $G$ being real, complex or quaternion real. 
Suppose also that the distribution of $G$ is unchanged by multiplication by a unitary matrix on the right.
We then have 
$$
A^\dagger A = U^\dagger G^\dagger G U \mathop{=}\limits^{\rm d} G^\dagger G = C,
$$
where the equality in distribution follows by the assumption of the appropriate invariance of the
distribution of $G$. From this, and noting $A$ is of size $N \times N$, the analogue of
(\ref{D1j}) reads
\begin{equation}\label{D2j}
(dA) \propto (\det C)^{(\beta/2) (-1+2/\beta)} (dC).
\end{equation}
Comparing (\ref{D1j}) and  (\ref{D2j}) we conclude
\begin{equation}\label{D3j}
(dG) \propto (\det A^\dagger A)^{(\beta/2)(n-N)} (dA).
\end{equation}
As noted in \cite{Fi12}, a corollary is that if the distribution of $G$ is 
\begin{equation}\label{gG}
g(G^\dagger G),
\end{equation}
with $G$ unitary invariant,
then $A$ has distribution proportional to
\begin{equation}\label{D4j}
(\det A^\dagger A)^{(\beta/2) (n - N)} g(A^\dagger A).
\end{equation}
And it follows from this that if the eigenvalue PDF for matrices $G$ distributed as $g(G^\dagger G)$ is given by $f(\lambda_1,\dots,\lambda_N)$,
then in the rectangular case with the construction (\ref{gA}), the eigenvalue PDF is proportional to
\begin{equation}\label{D4p}
\prod_{l=1}^N |z_l|^{\beta ( n - N)} f(z_1,\dots,z_N).
\end{equation}

The assumptions on $G$ leading to (\ref{D4j}), specifically the form (\ref{gG}) and its unitary invariance,
 hold for rectangular analogues of each of the Ginibre, spherical and truncated
unitary ensembles. Each has three sub-cases, depending on the elements being real $(\beta = 1)$, complex $(\beta = 2)$ or real  quaternion $(\beta = 4$).
In the Ginibre case, we take $G$ in (\ref{gA}) to be an $n \times N$ standard Gaussian matrix. The distribution on $G$ is then
proportional to $e^{-(\beta/2) {\rm Tr} \, G^\dagger G}$. In the spherical case we take $G = G_1^{-1} G_2$ where $G_1$ ($G_2$)
is an $n \times n$ $(n \times N)$ standard Gaussian matrix. The joint density of $G$ is then proportional to 
\cite[Ex.~3.6 q.3]{Fo10} 
\begin{equation}\label{2.46a}
\det( \mathbb I_N + G^\dagger G)^{-(\beta/2) ( n+N)}. 
\end{equation}
In the case of truncated unitary matrices
we begin with a unitary matrix of size $M \times M$ and then select a sub-block of size $n \times N$ which we call $G$.
For $M - (n+N) \ge 0$ we know \cite{Fo06} that the joint distribution of $G$ is proportional to 
\begin{equation}\label{2.40a}
\det (\mathbb I_N -
G^\dagger G)^{(\beta/2)(M- (n+N) + 1 - 2/\beta)}.
\end{equation}
Note that (\ref{2.46a}) and (\ref{2.40a}) reduce to (\ref{F1}) and (\ref{Aq}) with $\beta = 4$, after allowing for differences in notation.

In all these cases the eigenvalue PDF will have the structure (\ref{D4p}). Consider for definiteness the rectangular Ginibre
ensembles. The factor 
\begin{equation}\label{is}
\prod_{l=1}^N | z_l|^{\beta ( n - N)} = \prod_{l=1}^N e^{\beta ( n - N) \log | z_l|}
\end{equation}
 has the immediate
plasma interpretation of a repulsion from the origin by a charge of strength $(n-N)$. However this interpretation is not
informative from a random matrix viewpoint as is does not allow us to read off the eigenvalue density. Another interpretation
is therefore called for. We make use of Newton's theorem, which tells us that the potential created by a point charge is
the same as that created by a spherically symmetric continuous smeared out charge density when viewed from the outside of the latter.

Newton's theorem permits the interpretation of (\ref{is}) as due to a smeared out uniform charge density $1/\pi$ confined
to a radius $\sqrt{n-N}$ about the origin, with the plasma particles confined to lie outside this region. On the other hand,
we know that the factor $e^{-\beta | z_l|^2/2}$ results from a smeared out charge density $-1/\pi$ when viewed from inside
the plasma. These charge densities therefore cancel in the region $|z| < \sqrt{n-N}$. For charge neutrality the outer radius must
now extend to $\sqrt{n}$, so the total background charge density is equal to $-1/\pi$ in the annulus
\begin{equation}\label{is1}
\sqrt{n-N} < |z| < \sqrt{n}
\end{equation}
and is zero elsewhere. This particular plasma interpretation gives an immediate prediction for the eigenvalue density: to leading order
it will be uniform in the annulus (\ref{is1}) with density $1/\pi$. This is in keeping with the result from asymptotic analysis
of the analytic form for the eigenvalue density \cite{Fi12}, and provides an explicit example of the so-called single ring
theorem \cite{FZ97}.

The factor (\ref{is}) can be similarly interpreted in the spherical and truncated unitary ensembles, although there is an added
complication of $f$ in (\ref{D4p}) now also depending on $n$.
 For ease of presentation we
will restrict attention to the case of complex elements. Actually the plasma analogy for the complex spherical ensemble has
been discussed previously \cite{FF11}. 

The first point to note is that
substitution of (\ref{2.46a}) in (\ref{D4}) shows that for the spherical ensemble
the matrices $A$ as specified by 
(\ref{gA}) have distribution proportional to
$$
(\det A^\dagger A)^{(\beta/2)(n-N)}  \det( \mathbb I_N + A^\dagger A)^{-(\beta/2) ( n+N)}.
$$
In the case $\beta = 2$ (complex elements) this implies the joint eigenvalue PDF \cite[$N \mapsto n, M \mapsto N, n \mapsto N$]{FF11} 
proportional to 
\begin{equation}
\prod_{j=1}^N { |z_j|^{2(n-N)} \over (1 + |z_j|^2)^{n + 1}} \prod_{1 \le j < k \le N} | z_k - z_j|^2
\end{equation}
(note that this reduces to (\ref{CV}) in the case $n=N$).
Upon stereographic projection (\ref{SP}) this takes on the form
\begin{equation}\label{CV1b}
\prod_{l=1}^N |\vec{r}_l|^{2(n-N)} \prod_{1 \le j < k \le N} |\vec{r}_k - \vec{r}_j|^2,
\end{equation}
where as in (\ref{CV1}) 
$\vec{r}_j$ is the vector in $\mathbb R^3$ corresponding to the point $(\theta_j,\phi_j)$ on the sphere, with the origin at the north pole.

We seek a one-component plasma system on the sphere with a uniform background charge that has Boltzmann factor
(\ref{CV1b}). An important point is that on the sphere each charge is accompanied by a neutralising smeared out
uniform charge density (recall (\ref{SSP})). This means that if we consider a spherical cap about the north pole specified by
$0 < \theta < \theta^*$, of area $A_{\theta^*} = {\pi \over 2} (1 - \cos \theta^*)$, then the uniform charge density in that region is $-N A_{\theta^*}/\pi$.
We want to impose an external uniform charge density $A_{\theta^*}/\pi$ in $\theta^* < \theta < \pi$, which
furthermore will be the region occupied by the mobile charges. Doing this, and choosing
\begin{equation}\label{2.43a}
{A_{\theta^*} \over \pi} = {n/N - 1 \over n/N},
\end{equation}
it is shown in \cite{FF11} that (\ref{CV1b}) is, up to a multiplicative constant, the Boltzmann factor for the plasma system.
Note that this predicts that the eigenvalue density will to leading order be supported on the portion of the sphere specified by
the azimuthal angle being in the range $\theta^* < \theta < \pi$, and will be uniform. We remark too that in the complex
plane, the formula (\ref{2.43a}) and the stereographic projection formula (\ref{SP}) tell us that the inner boundary
of support $r^*$ say satisfies $(r^*)^2 = {n \over N} - 1$ \cite{FF11}, while the density profile in this support is
\begin{equation}\label{2.43b}
{n \over \pi (1 + |z|^2)}.
\end{equation}

It remains to identify the plasma system for rectangular truncated unitary matrices subject to the construction (\ref{gA}).
Here, from the discussion in the paragraph below (\ref{D4p}), and
(\ref{D4}), the matrices $A$ as specified by (\ref{gA}) have distribution proportional to
$$
(\det A^\dagger A)^{(\beta/2)(n-N)} 
\det (\mathbb I_N -
A^\dagger A)^{(\beta/2)(M- (n+N) + 1 - 2/\beta)}.
$$
The corresponding joint eigenvalue PDF is known to be proportional to \cite{Fi12}
\begin{equation}\label{sth}
\prod_{j=1}^N
\Big ( { | z_j |^{2}  \over 1 - |z_j|^2 } \Big )^{n-N}
(1 - | z_j|^2)^{M-N-1} \prod_{1 \le j < k \le N} | z_k - z_j|^2
\end{equation}
(note that this gives (\ref{D1}) in the case $n=N$ and $M \mapsto N + n$).
To relate this to the Boltzmann factor for a plasma system, we proceed as for the task of similarly interpreting (\ref{D1}).

First, a smeared out neutralising charge density $-(\eta + N/4 \pi a^2)$ is to be imposed on the psuedosphere,
and we set $a=1/2$.
We know that for the corresponding one-component plasma system (\ref{D3}) is proportional to the Boltzmann factor.
Next, for $\tau > \tau_-$ the potential, by Newton's theorem, will be that of a charge at the origin of strength
\begin{equation}\label{QQ}
Q = A_{\tau_-} (\eta + N/4 \pi a^2), 
\end{equation}
where $A_{\tau_-} = 2 \pi a^2(\cosh \tau_- - 1)$  is the area of the region $0 < \tau < \tau_-$.
According to (\ref{D2a}), this potential is therefore equal to $-Q \log (|z|/(1 - |z|^2)^{1/2}$, and so with the particles confined to
the region $\tau > \tau_-$ the Boltzmann factor is given by (\ref{D3}) multiplied by
$\prod_{j=1}^N (|z_j|^2/(1 - |z_j|^2))^Q$, giving us in total the functional form (\ref{sth}) with 
\begin{equation}\label{QQ1}
Q = n - N
\end{equation}
 and
$\pi \eta + 1 = M - N - 1$.  
Let us suppose $R_-$ is related to $\tau_-$ by stereographic projection. 
Noting from (\ref{SS}) that the area on the pseudosphere with $a=1/2$, corresponding to a disk of radius $R_-$ 
in the complex plane centred on the origin,
is equal to $\pi  R_-^2/ (1 - R_-^2)$, we then we have from (\ref{QQ}) that
$Q = (M-2) R_-^2 /(1 - R_-^2)$, and this equated with (\ref{QQ1}) tells us that for large $n,M,N$,
\begin{equation}\label{T1}
R_-^2 = {(n-N)/M \over 1 + (n-N)/M}.
\end{equation}

The outer radius $R_+$ of the support of the eigenvalue density can similarly be computed. Since the background
charge in $|z| < R_-$ is $-Q = - (n - N)$ and this has been neutralised, we require that the total charge
for $|z| < R_+$ be equal to $N+Q = n$ so has to have total charge in $R_- < |z| < R_+$ equal to $N$.
Recalling too from below (\ref{QQ1}) that for large $M-N$, $\pi \eta \approx M - n$ we obtain
\begin{equation}\label{T2}
R_+^2 = {1 \over 1 + (M-N)/n}.
\end{equation}
Note that this is consistent with (\ref{RR}) once we make the identification noted below (\ref{sth}).
These formulas are in agreement with the exact results reported in \cite{Fi12}.

\subsection{Products of random matrices}\label{S2.6}
Consider the product of $m$ independent Ginibre matrices.
Considerations from free probability \cite{BJW10}, \cite{GT10}, \cite{OS11} tell us that
\begin{equation}\label{L3eA}
\lim_{N \to \infty} {1 \over N} N^m  \rho_{(1)}(N^{m/2}w) = {|w|^{(2/m) - 2} \over m \pi} \chi( |w|<1).
\end{equation}
Let us suppose the Ginibre matrices have complex elements. A recent advance in random matrix theory
has been the exact determination of the joint eigenvalue PDF \cite{AB12}, which has been shown to be
proportional to 
\begin{equation}\label{2ka}
\prod_{l=1}^N w_m^{(2)}(|z_l|) \prod_{1 \le j < k \le N} |z_k - z_j|^2, \qquad
w_m^{(2)}(|z|) = 
G^{m,0}_{0,m}\Big ({ \underline{\hspace{0.5cm}} \atop 0, \dots, 0} \Big | |z|^2 \Big ).
\end{equation}
Here $G^{m,0}_{0,m}$ is an example of the Meijer G-function; see e.g.~\cite{Lu69}. From a plasma viewpoint,
we can replace $w^{(2)}(|z|)$ by its large $|z|$ form, 
\begin{equation}\label{ZL}
w_m^{(2)}(|z|) \mathop{\sim}\limits_{|z| \to \infty} e^{- m |z|^{2/m} + O(\log |z|)}
\end{equation}
obtained from known asymptotics of the Meijer G-function \cite{Lu69}. Substituting this in (\ref{2ka}),
and going through the argument below (\ref{2.ta}), we see that the plasma viewpoint correctly predicts 
(\ref{L3eA}).

Note that changing variables $|w|=r^m$ in (\ref{L3eA}) reclaims the circle law. An understanding of this
observation has been given in \cite{BNS12}, as a corollary of the fact that the limiting global spectral
density of a product of $m$ Ginibre matrices is equal to that for a single Ginibre matrix raised to the $m$-th
power. The mechanism for this --- that Ginibre matrices are unchanged in distribution by multiplication on the
left or right by unitary matrices in the same number field --- also holds true for the spherical ensemble and
truncations of unitary matrices. 

Thus for a product of $m$ independent $N \times N$ matrices from the spherical ensemble, changing variables
$|z| = |w|^{1/m}$ in the stereographic projection of the uniform density on the sphere of radius $1/2$,
$$
{1 \over \pi (1 + |z|^2)^2}
$$
(cf.~(\ref{2.43b}))
gives the asymptotic density \cite{GKT14}
\begin{equation}\label{nd1}
{N |w|^{2/m - 2} \over \pi m (1 + |w|^{2/m})^2}.
\end{equation}
Similarly, the same change of variables in (\ref{RR}) gives, for the product of $m$ independent truncations
of ${\rm U}(n+N)$ matrices of size $N \times N$, the asymptotic density \cite{BNS12}
\begin{equation}\label{nd2}
{(n+N) w^{2/m - 2} \over \pi m (1 - |w|^{2/m})^2} \chi_{|w| < R^m},
\end{equation}
where $R$ is given in (\ref{RR}).

To understand (\ref{nd1}) and (\ref{nd2}) from a plasma viewpoint, we first note from \cite{ARRS13} and
\cite{ABKN14} (see also the review \cite{AI15}) that in the case of complex entries the eigenvalue PDF
for products of $m$ spherical ensemble matrices, or  $m$ truncated unitary matrices, is again of
the form (\ref{2ka}), but with $w_m(|z|)$ replaced by
\begin{equation}\label{Gh1}
G^{m,m}_{m,m}\Big ({-N,\dots,-N \atop 0, \dots, 0} \Big | |z|^2 \Big ), \qquad
G^{m,0}_{m,m}\Big ({n,\dots,n \atop 0, \dots, 0} \Big | |z|^2 \Big ).
\end{equation}
From the differential equation satisfied by the Meijer $G$-function \cite{Lu69}, we have that for large $N,n$ and
with $|z|^2 = x$ these functions satisfy the differential equations
\begin{equation}
\Big ( (-1)^m x \prod_{j=1}^m ( x {d \over dx} + N ) - \prod_{j=1}^m (x {d \over dx} )\Big ) G = 0, \qquad
\Big ( x \prod_{j=1}^m ( x {d \over dx} - n ) - \prod_{j=1}^m (x {d \over dx}) \Big ) G = 0.
\end{equation}
respectively. 
Being of order $m$, each admits $m$ linearly independent solutions. Of interest to us are the particular large $n,N$ asymptotic solutions
\begin{equation}\label{ZL1}
G = {1 \over (1 + m x^{1/m})^N}, \qquad G = (1 - m x^{1/m})^n.
\end{equation}
We remark that asymptotic properties of the Meijer G-function differential equation is also a feature of the recent
studies \cite{FL15,FW15} relating to the singular values of certain products of random matrices.
We now use (\ref{ZL1})  in place of the Meijer G-functions in (\ref{Gh1}), and thus in place of $w_m$ in (\ref{2ka}).
A plasma derivation of (\ref{nd1}) and (\ref{nd2}) can now be given by going through the argument below (\ref{2.ta}).

The above discussion applies when the matrices in the products have complex entries. Products from the same
ensembles but with real quaternion entries allows for a similar analysis. This is because the
eigenvalue PDF maintains the same structure as for the $m=1$ cases, but with a different
one-body weight \cite{Ip13}, \cite{AI15},
$$
\prod_{l=1}^N w_m^{(4)}(|z_l|) | z_l - \bar{z}_l |^2 \prod_{1 \le j < k \le N} |z_k - z_j|^2 |z_k - \bar{z}_j|^2.
$$
Moreover, the weight $w_m^{(4)}$ is simply related to $w_m^{(2)}$ appearing in (\ref{2ka}) \cite[Eq.~(2.46)]{AI15},
allowing for its replacement by the asymptotic forms in (\ref{ZL}) and (\ref{ZL1}) as appropriate. In particular, this tells
us that the asymptotic densities will be those known for the $m=1$ case, but with the change of variables $|z| = |w|^{1/m}$.

In the case of products of random matrices from the Ginibre, spherical or truncated unitary matrices with real
entries, the structure of the joint eigenvalue PDF (which we recall breaks up into sectors due to the real eigenvalues)
again is the same as for the $m=1$ cases but with different weights \cite{IK14}. However now the weight for
the complex conjugate pairs does not have a closed form beyond the cases $m=1$ and
$m=2$ \cite{Ed97,APS10}, so the task of obtaining its asymptotic form remains open.

\section{Sum rules}\label{S3}

Coulomb systems in general, and the two-dimensional one-component plasma in particular, exhibit a number features
characteristic of the long range nature of the pair potential. Many of these features show themselves by way of
sum rules for the corresponding correlation functions. By way of example, let $\rho_{(2)}(\vec{r}, \vec{0})$
denote the two-particle correlation function in the bulk of the two-dimensional one-component plasma, and let
$\rho_{(2)}^T(\vec{r}, \vec{0}) := \rho_{(2)}(\vec{r}, \vec{0}) - \rho^2$ denote the truncated (or connected) two-particle
correlation. For the plasma at inverse temperature $\beta$, these satisfy the sequence of sum rules for the moments
\cite{Ma88,KMST99}
\begin{align}
&{1 \over \rho} \int_{\mathbb R^2} \rho_{(2)}^T(\vec{r},\vec{0}) \, d \vec{r} = - 1 \label{1a}\\
& \int_{\mathbb R^2} r^2 \rho_{(2)}^T(\vec{r},\vec{0}) \, d \vec{r} = - {2 \over \pi \beta}  \label{1b}\\
& \rho  \int_{\mathbb R^2} r^4 \rho_{(2)}^T(\vec{r},\vec{0}) \, d \vec{r} = - {16 \over (\pi \beta)^2} \Big ( 1 - {\beta \over 4} \Big )  \label{1c}\\
&\rho^2  \int_{\mathbb R^2} r^6 \rho_{(2)}^T(\vec{r},\vec{0}) \, d \vec{r} = - {18 \over (\pi \beta)^3} \Big ( \beta -  6 \Big ) \Big ( \beta - {8 \over 3}  \Big ).   \label{1d}
\end{align}
The first of these has the interpretation that upon a charge being fixed at the origin, the system responds by creating an image
charge of equal and opposite total charge; thus rewrite this to read
$$
 \int_{\mathbb R^2} \Big ( \rho_{(2)}^T(\vec{r},\vec{0})  + \rho \delta(\vec{r}) \Big )\, d \vec{r} = 0 .
 $$
 The second is known as the Stillinger-Lovett sum rule, and can be derived as a consequence of the physical requirement
 that the plasma perfectly screens an external charge density in the long wavelength limit (see e.g.~\cite[\S 15.4.1]{Fo10}).
 The third is known as the compressibility sum rule, and can be viewed as a refinement of the linear response derivation
 of (\ref{1b}) (see e.g.~\cite[\S 14.1.1]{Fo10}). The sixth moment sum rule (\ref{1d}), first derived in \cite{KMST99} has recently been
 shown to result as a consequence of the response to variations in spatial geometry \cite{CLW14a,CLW14b}.
 
 The bulk truncated two-point correlation function for the complex Ginibre ensemble (which is also shared by all the other ensembles
 considered in Section \ref{S2}, in an appropriate scaling and for regions that the density is constant) has the explicit form
\begin{equation}\label{T3} 
 \rho_{(2)}^T(\vec{r},\vec{0}) = - \rho^2 e^{- \pi \rho |\vec{r}|^2},
 \end{equation}
 with $\rho = 1/\pi$ (see e.g.~\cite[\S 15.3.2]{Fo10}). The sum rules (\ref{1a})--(\ref{1d}) with $\beta = 2$ are readily verified.
 
There are also sum rules which contrast the behaviour of the plasma at the boundary to its bulk behaviour.
Consider for example  the asymptotic form of the charge-charge
correlation in the vicinity of the boundary, taken to be $y=0$ in the scaled limit for definiteness. This quantity is
predicted \cite{Ja82} to have a slow decay in the direction of the boundary according to an explicit 
one on distance squared decay, which in the case of the one-component plasma implies the sum rule for the
truncated correlation
\begin{equation}\label{T3} 
\rho_{(2)}^T((x_1,y_1),(x_2,y_2)) \mathop{\sim}\limits_{|x_1 - x_2| \to \infty}
- {f(y_1,y_2) \over 2 \beta \pi^2 ( x_1 - x_2)^2}, \qquad {\rm with} \quad \int_{-\infty}^\infty \int_{-\infty}^\infty f(y_1,y_2) \, dy_1 dy_2  = 1.
\end{equation}

For the complex Ginibre ensemble, the scaled edge correlations $\rho_{(k)}^{\rm edge}$ are defined by
$$
\rho_{(k)}^{\rm edge}((x_1,y_1),\dots,(x_k,y_k)) = \lim_{N \to \infty} \rho_{(k)}((x_1,-\sqrt{N}+y_1),\dots,(x_k,-\sqrt{N}+y_k)),
$$
and explicit computation of this limit gives \cite{FH98}, \cite[Prop.~15.3.5]{Fo10}
\begin{equation}\label{T4} 
\rho_{(2)}^{\rm edge}((x_1,y_1),(x_2,y_2)) =  \Big (
H(y_1) H(y_2) - e^{- (x_1 - x_2)^2 - (y_1 - y_2)^2}
\Big | H \Big ( {1 \over 2} (y_1 + y_2 + i (x_1 - x_2)) \Big ) \Big |^2 \Big ),
\end{equation}
where 
\begin{equation}\label{HH}
H(z) = {1 \over 2 \pi} \Big (1 + {\rm erf} (\sqrt{2} z) \Big ). 
\end{equation}
One calculates from this the large $|x_2 - x_1|$ asymptotic
form
\begin{equation}\label{T5} 
\rho_{(2)}^{T \, {\rm edge}}((x_1,y_1),(x_2,y_2)) \mathop{\sim}\limits_{|x_1 - x_2| \to \infty} - {H'(y_1) H'(y_2) \over 4 \pi^2 (x_1 - x_2)^2} , 
\end{equation}
which indeed complies with (\ref{T3}) in the case $\beta = 2$, as noted in \cite{FH98}.

The physical origin of the slow decay is the nonzero dipole moment of the screening cloud, due to the presence of the
spectral boundary \cite{Ja82}. This can be quantified from the viewpoint of the density profile.
 With $R$ denoting the boundary of the neutralising background
of charge density $-\rho_b$, define the boundary density profile $\rho_{(1)}^{\rm edge}(r) = \lim_{N,R \to \infty} \rho_{(1)}(R-r)$.
Results in \cite{TF12,Wi12} give the dipole moment sum rule
\begin{equation}\label{HHs}
\int_{-\infty}^\infty r ( \rho_{(1)}^{\rm edge}(r) -  \chi_{ r>0} \rho_b) \, dr = - {1 \over 2 \pi \beta} (1 - \beta/4).
\end{equation}
Note that the $\beta = 2$ density profile (\ref{HH}) complies with this sum rule.

With the moments of the screening cloud near a boundary in mind, let us consider the limiting truncated two-point correlation in
the neighbourhood of the $x$-axis for the real quaternion Ginibre ensemble. This has been calculated in 
\cite{Ka98}, \cite[Ex.~15.9 q.2]{Fo10} to be given by
\begin{equation}\label{T5} 
\rho_{(2)}^{T \, {\rm rqG}} ((x_1,y_1),(x_2,y_2))  = - 4 y_1 y_2 e^{-(x_1^2 + y_1^2) - (x_2^2+y_2^2)} \Big (
f(z_1,\bar{z}_2) f(z_2,\bar{z}_1) + f(z_1,z_2) f(\bar{z}_1, \bar{z}_2) \Big ),
\end{equation}
where $z_1 = x_1 + iy_1$, $z_2 = x_2 + i y_2$ and
\begin{equation}\label{T5f} 
f(w,z) = {i \over \sqrt{2 \pi}} e^{(w^2 + z^2)/2} {\rm erf} \, \Big ( {z - w \over \sqrt{2}} \Big ).
\end{equation}
In contrast to the algebraic decay (\ref{T4}) of the correlations along the boundary in the complex Ginibre ensemble,
the correlation (\ref{T5}) decays as a Gaussian in the direction of the $x$-axis, which for the real quaternion
Ginibre ensemble is the boundary between the eigenvalues in the upper half plane, and their complex conjugate
pairs. 

Converse to the slow decay (\ref{T4}) implying a non-zero dipole moment of the screening cloud, a fast
decay implies that all complex (multi-pole) moments of the screening cloud much vanish (see e.g.~\cite{Ma88}).
As already noted, the complex conjugate eigenvalues behave like image charges, and we know from \cite{Ja82,Fo85}
that in this circumstance we must interpret the screening cloud as the function of $(x_2,y_2) \in \mathbb R^2$
specified by
\begin{equation}\label{T6} 
\rho_{(2)}^T((x_1,y_1),(x_2,y_2))  + \delta(x_1-x_2) \delta(y_1-y_2) \rho_{(1)}((x_2,y_2)).
\end{equation}
Here $\rho_{(2)}^T((x_1,y_1),(x_2,y_2)) $ is given by (\ref{T5}), which we note is even in $y_2$.
The quantity $ \rho_{(1)}((x_2,y_2))$ is the limiting eigenvalue density for the real quaternion
ensemble, which we know from \cite{Ka98}, \cite[Ex.~15.9 q.2]{Fo10} to be given by
\begin{equation}\label{T6a} 
\rho_{(1)}^{{\rm rqG}} ((x,y)) = 2 y e^{-(x^2 + y^2)} f(z,\bar{z}),
\end{equation}
and is also even in $y_2$.

By translation invariance of the system along the $x$-axis we can set $z = i y_1$. Due to (\ref{T6})
being even in $y_2$ the odd complex moments  vanish by symmetry, while the vanishing of the even complex
moments requires
\begin{equation}\label{T7} 
\int_{\mathbb R \times \mathbb R_+} w^{2p} \rho_{(2)}^T(z,w) \, d^2 w = - z^{2p} \rho_{(1)}(z), \qquad p=0,1,\dots
\end{equation}
If we multiply both sides by $\alpha^p/p!$ and sum over $p=0,1,\dots$, then substitute (\ref{T6}) and (\ref{T6a}),
we obtain the equivalent form
\begin{equation}\label{T8} 
    \int_{\mathbb R^2} y e^{\alpha w^2} 
 e^{ - |w|^2} \Big (
f(iy_1,\bar{w}) f(w, - i y_1) + f(iy_1,w) f(-i y_1, \bar{w}) \Big )
\, d^2 w =   e^{-\alpha y_1^2}   f(iy_1,-i y_1),
\end{equation}
where $w=x+iy$.
To verify this, we substitute (\ref{T5f}) on the LHS, change variables  $x \mapsto x - i y$, write
$y e^{-2 y^2} = -{1 \over 4}{d \over dy}  e^{-2y^2}$, and integrate by parts with respect to $y$.
The integrand then consists of two terms. Integrating over $y$ in the first of these gives
$\pi \delta(x+i y_1)$, and integrating over $y$ in the second gives $pi \delta(x-i y_1)$. Thus
in both cases the integrations over $x$ are immediate, and the RHS results.

The correlations in the neighbourhood of the real axis for the real Ginibre ensemble \cite{FN07,BS09} can be analysed from
an analogous viewpoint. Thus they decay at the rate of a Gaussian, and so Coulomb gas theory predicts that the screening
cloud, extended to include image charges, must have the property that all complex moments vanish. Due to the
presence of a finite density of real eigenvalues, the screening cloud will involve both the correlation between two eigenvalues
in the upper half plane, and an eigenvalue on the real axis (see \cite[Eq.~(71)]{FM11}. The verification of this sum rule from the
explicit functional form is known from \cite[Prop.~4.8]{FM11}.

The plasma analogy valid for the products of random matrices considered in
Section \ref{S2.6} tells us that it must also be that the complex moments of the screening cloud for
the limiting correlation functions in the neighbourhood of the real axis vanishes, although we don't
undertake a verification here.

\section{Random self-dual non-Hermitian matrices}\label{S4}
Dyson's three fold way relates  Hermitian matrices  ${1 \over 2} (G + G^\dagger)$ corresponding to square
Ginibre matrices $G$ to global time reversal symmetries \cite{Dy62c}. 
There are similarly three classes of chiral ensembles corresponding to the  global time reversal symmetry of the Dirac 
Hamiltonian; see e.g.~\cite{Ve94}.
This viewpoint was extended by Atland and
Zirnbauer \cite{AZ97}, who introduced a further four random matrix ensembles motivated by the theoretical description
of conductance through a normal metal -- superconductor junction in terms of a matrix Bogoliubov -- de Gennes Hamiltonian
$$
\mathcal H = \begin{bmatrix} h & \Delta \\ - \bar{\Delta} & - h^T \end{bmatrix}, \qquad \Delta = - \Delta^T.
$$
With each block in this matrix of size $N \times N$, $N$ even, the appropriate time reversal symmetry operator $T$ is such that
$T^2 = -1$ and has the special structure 
$$
T = \mathbb I_2 \otimes \begin{bmatrix} \mathbb O_{N/2} & \mathbb I_{N/2} \\
- \mathbb I_{N/2} & \mathbb O_{N/2} \end{bmatrix} K,
$$
where $K$ denotes complex conjugation. The requirement that $\mathcal H$ commute with $T$ shows, after rearrangement of rows
and columns in the former, that $\mathcal H$ has the block form
$$
\mathcal H = \begin{bmatrix}  A & B \\ - \bar{B} & - \bar{A} \end{bmatrix}
$$
with $A = A^\dagger$ and $B = - B^T$, and furthermore the elements of $A$ and $B$ real quaternions. As shown in 
e.g.~\cite[Ex.~3.3 q.1]{Fo10}, the eigenvalue problem for this structure is equivalent to the eigenvalue problem for 
\begin{equation}\label{T8a} 
\begin{bmatrix} \mathbb O_N & W \\ W^\dagger & \mathbb O_N \end{bmatrix},
\end{equation}
where $W$ is an $N \times N$ antisymmetric matrix with complex elements, which in turn is equivalent to the eigenvalue
problem for
\begin{equation}\label{T8ba} 
\begin{bmatrix} \mathbb O_N & D \\ D^\dagger &  \mathbb O_N \end{bmatrix},
\end{equation}
where $D$ is an $N \times N$ self-dual matrix --- meaning that $D = Z_N D^T Z_N^{-1}$, where
$Z_N = \mathbb I_{N/2} \otimes \begin{bmatrix} 0 & - 1 \\ 1 & 0 \end{bmatrix}$ --- with complex elements.  On this last
point, one can check that if $W$ is antisymmetric, then $Z_N W$ is self dual. A feature of self dual matrices is that
their eigenvalues are doubly degenerate.

The structures (\ref{T8a}) and (\ref{T8ba}) motivated Hastings \cite{Ha01} to investigate the eigenvalue distribution on
the non-Hermitian self dual matrix $D$, constructed according to $D=Z_N W$, where $W$ is a  anti-symmetric
matrix with all independent entries  complex Gaussians with zero mean and fixed standard deviation $\sigma$ say for
both the real and imaginary parts. In contrast to the ensembles of non-Hermitian matrices discussed in Section \ref{S2},
the eigenvalue PDF for the ensemble of self-dual non-Hermitian complex Gaussian matrices is not known explicitly.
Nonetheless, some analysis has been possible under the assumption that the eigenvalues are at large separation,
suggesting that in this limit the eigenvalue PDF is well approximated by \cite{Ha01}
\begin{equation}\label{T8b} 
\prod_{l=1}^N e^{- {1 \over 2 \sigma^2} | z_l |^2} \prod_{1 \le j < k \le N} | z_k - z_j|^4,
\end{equation}
up to proportionality. Writing this in the Boltzmann factor form $e^{-\beta U}$ gives that $U$ is given by
(\ref{U1}), except that the coefficient of the first term is ${1 \over 8 \sigma^2}$, and in distinction
to (\ref{Z1}) the inverse temperature is now $\beta = 4$.

Continuing the investigations already initiated in \cite{Ha01}, one would like to show that the properties of the functional
form (\ref{T8b}) are consistent with observed properties of the eigenvalues. The most immediate, already noted in
\cite{Ha01}, is that in the plasma picture the neutralising background charge density $- \rho_b$ say responsible
for the harmonic term in (\ref{T8b}) must satisfy the Poisson equation $\nabla^2 ( {1 \over 8 \sigma^2} | z|^2 ) = 2 \pi \rho_b$,
and so $\rho_b = {1 \over 4 \pi \sigma^2}$. With $N$ eigenvalues the leading spectral radius is thus predicted to be
$2 \sigma \sqrt{N}$ and the density to be uniform inside this region --- these features are clearly evidenced by a plot of computer
generated eigenvalues for matrices from the ensemble (see e.g.~\cite[Fig.~1. Here $\sigma = 1/\sqrt{2}$]{Ha01}).

\begin{figure}[!t]
%%%%%%%%%%%%%%%%%%%%
\begin{center}
\includegraphics[scale=0.5]{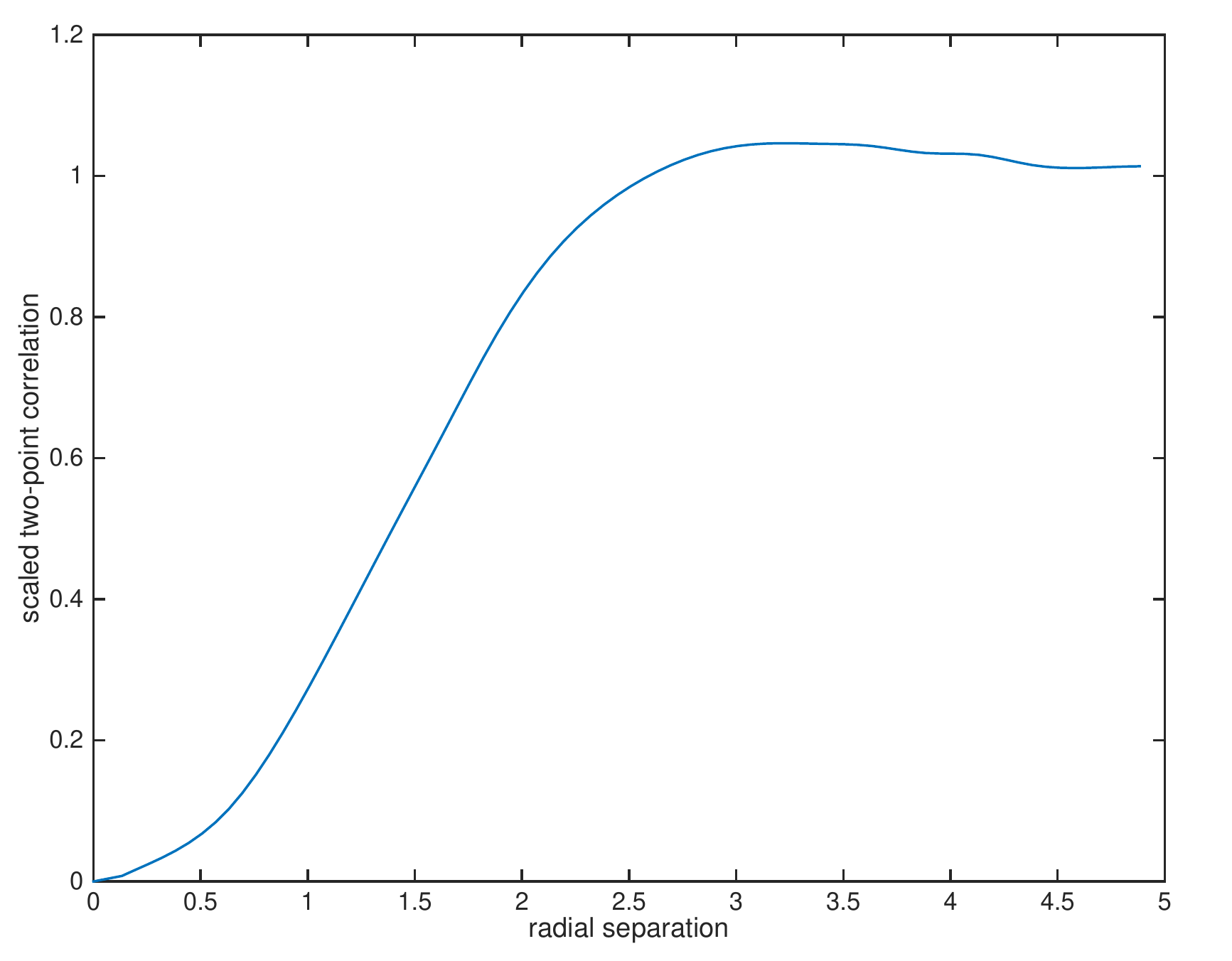}
\end{center}
\caption{Plot of $\rho_{(2)}^T(\vec{r},\vec{0})/\rho_b^2$, $\rho_b = 1/2\pi$, as calculated from the simulation of
$10^6$ non-Hermitian self dual matrix with independent entries standard complex Gaussian entries of size $140 \times 140$,
and smoothed using a cubic spline. \label{Fs1a}}
\end{figure}

The main focus of attention in \cite{Ha01} is the one-point density profile and the two-point correlation function.
Regarding the latter, in the plasma system one sees from the explicit functional form (\ref{T3}) that for $\beta = 2$ the
truncated correlation function is always negative and furthermore is monotonically increasing from its value $-\rho^2$
at $|\vec{r}| = 0$ to its value zero as $|\vec{r}| \to \infty$. This is consistent with all the moments (\ref{1a})--(\ref{1d})
being negative when $\beta = 2$. In contrast, at $\beta = 4$ we see from (\ref{1c}) the fourth moment vanishes while
(\ref{1d}) gives that the 6th moment is positive. Hence, for $\beta = 4$ in the plasma system the truncated two-point
correlation function must become positive and in particular is not monotonic. This behaviour is clearly seen in
computer generated plots for finite size systems --- see e.g.~\cite[Fig.~2]{TF99} --- which exhibit a peak in
$\rho_{(2)}^T(\vec{r},\vec{0})$ at $|\vec{r}| \approx 1.7/(\pi \rho_b)^{1/2}$. On the random matrix side, the data presented
in \cite{Ha01} was inconclusive due to statistical errors. However, this latter restriction is readily overcome with minimal
effort due to the advances in desktop computing (we used Matlab\_R2014b). The results, displayed in Figure \ref{Fs1a}
in the case $\rho_b = 1/(2 \pi)$, provide conclusive evidence for a peak in the truncated two-point correlation function at 
$|\vec{r}| \approx 3$, in agreement with the above quoted peak for the plasma system, although the exact profiles are
different. Most evident is that $\rho_{(2)}^T(\vec{r},\vec{0})$ is proportional to $|\vec{r}|^4$ for small $\vec{r}$ in the plasma system, whereas
it appears to vanish in proportion to $|\vec{r}|^2$ for the random matrix ensemble. Note that this latter point does not contradict
(\ref{T8b}), which is only predicted to be accurate for large separations. Although our simulations are accurate at a graphical
level, they do not provide reliable estimates of the moments, so the question as to whether the sum rules (\ref{1a})--(\ref{1d})
with $\beta = 4$ are valid for the random matrix ensemble remains open.

\begin{figure}[!t]
%%%%%%%%%%%%%%%%%%%%
\begin{center}
\includegraphics[scale=0.5]{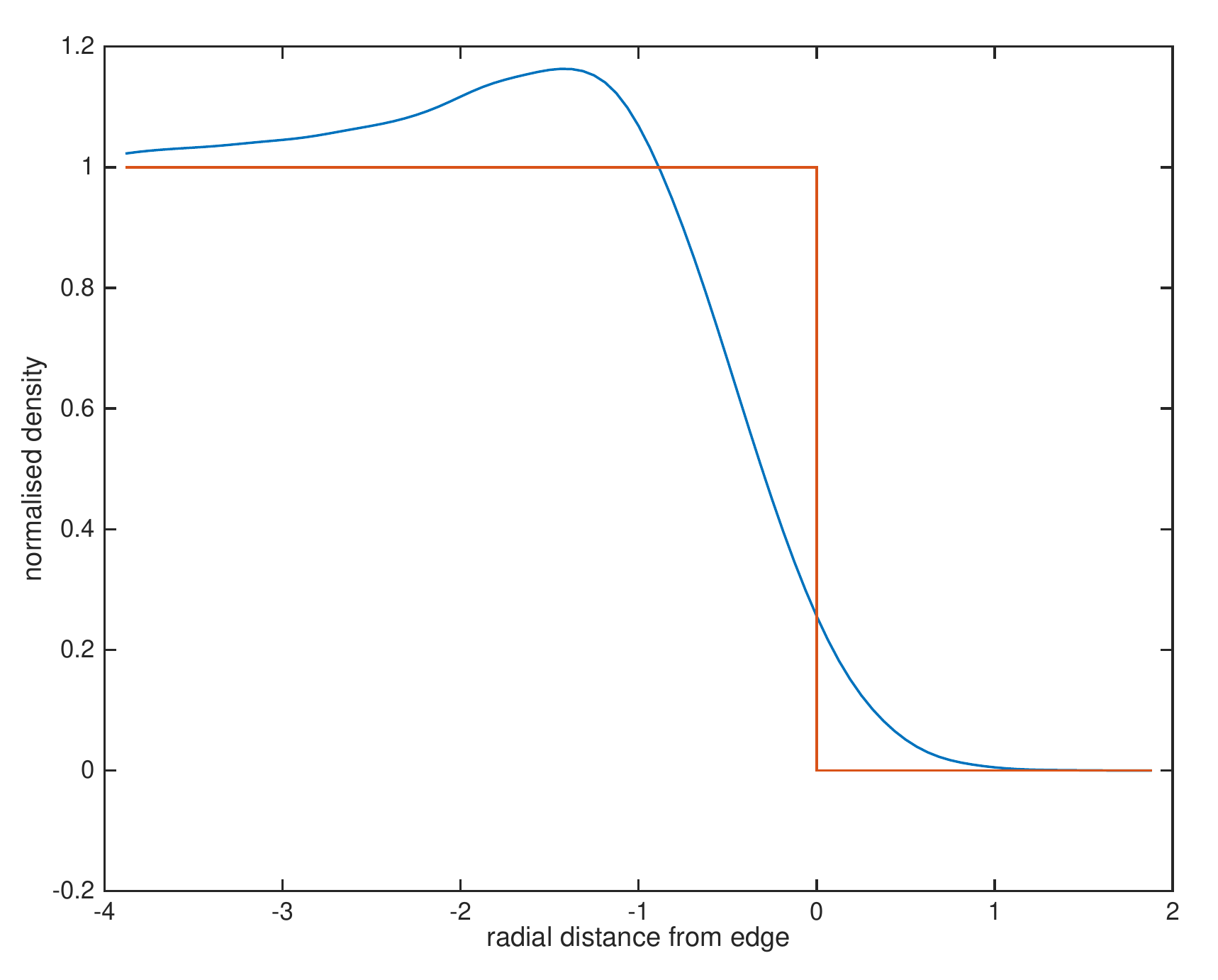}
\end{center}
\caption{Plot of $\rho_{(1)}(\vec{r})/\rho_b$, $\rho_b = 1/2\pi$, as calculated from the simulation of
$10^6$ non-Hermitian self dual matrix with independent entries standard complex Gaussian entries of size $140 \times 140$,
and smoothed using a cubic spline.
The origin has been shifted to $\sqrt{2N}$, which with $\sigma = 1/\sqrt{2}$ is the predicted spectral radius, and as a visual
aid a step function density profile inside the spectral radius has been superimposed. \label{Fs2a}
}
\end{figure}

Next we turn our attention to the radial density. Results from our simulation are displayed in Figure \ref{Fs2a}.
As already noticed in \cite{Ha01}, this profile is dramatically different to the scaled edge density $H(r)$ as given by (\ref{HH}) for
complex Ginibre matrices, or equivalently the plasma system with $\beta = 2$, which does not display any overshoot
phenomenon, i.e.~ a peak in the spectral density near the spectral edge.  In the work \cite{CFTW14} the overshoot effect observed
for the density profile in the non-Hermitian self dual matrix  is predicted to be a feature of the two-dimensional one-component plasma
system density profile for all $\beta \ge 2$, and the validity of this prediction was given analytic confirmation by a perturbation expansion
about $\beta = 2$ in \cite{CFTW15}. Note that an overshoot must happen for all $\beta \ge 4$ at least
to be consistent with the sum rule (\ref{HHs}).
Comparison of Figure \ref{Fs2a} with Figure 1, label `2'  in  \cite{CFTW14}, shows quantitative
agreement, although there are differences at a qualitative level, for example in Figure \ref{Fs2a} the density profile does not dip below
the asymptotic value in the direction of the bulk, but does in Figure 1, label `2'  in  \cite{CFTW14}.

In summary, our investigation adds to the
evidence first presented in \cite{Ha01} that eigenvalues from the ensemble of non-Hermitian self dual matrices have
global behaviours consistent with those of the plasma system (\ref{T8b}), although the precise 
functional forms of the correlations are different. Due to this latter fact, it is not known if the sum rules
(\ref{HHs}) and (\ref{1a})--(\ref{1b})
with $\beta = 4$, characteristic of the plasma system, remain valid for the random matrix ensemble.

\section{Plasma analogy for eigenvalues of a single row and column truncation of CSE matrices}\label{S5}
In Section \ref{S2.3} truncated unitary matrices were discussed from the viewpoint of plasma analogies for the
corresponding eigenvalue PDFs. The unitary matrices considered were from the three classical groups ${\rm U}(n+N)$, ${\rm Sp}(2(n+N))$
and ${\rm O}(n+N)$ with the uniform (Haar) measure. Very recently \cite{KK15} the explicit eigenvalue PDF for a 
single row and column truncation of Dyson's three circular ensembles, the COE $(\beta = 1$), CUE $(\beta = 2$) and CSE
$(\beta = 4$) has been calculated as
\begin{equation}\label{1b2} 
{\beta^N \over (2 \pi)^N}
\prod_{j,k=1}^N (1 - z_j \bar{z}_k)^{\beta/2 - 1} \prod_{1 \le j < k \le N} | z_k - z_j |^2.
\end{equation}
This PDF is supported on $|z_l |< 1$, $(l=1,\dots,N)$. 

One recalls (see e.g.~\cite[Ch.~2]{Fo10}) that the CUE is made up of matrices from ${\rm U}(N)$ with Haar measure which thus
explains why (\ref{1b2}) with $\beta = 2$ coincides with (\ref{D1}) with $n=1$. On the other hand matrices from the COE and
CSE are not the same as matrices from  ${\rm O}(N)$ and ${\rm Sp}(2N)$. Matrices from the COE are constructed from
matrices from $U \in {\rm U}(N)$ by forming $U U^T$, while matrices from the CSE are constructed from $U \in {\rm U}(2N)$ by forming
the self dual quaternion matrices $Z_{2N}^{-1} U^T Z_{2N} U$, where $Z_{2N}$ is defined below (\ref{T8b}). For the truncations considered
in \cite{KK15}, $N$ is replaced by $N+1$, and in the case of the CSE a truncation refers to a row and column of real quaternion elements.
As with the original CSE matrices, the resulting matrices again have doubly degenerate eigenvalues.

Of the cases $\beta = 1$ and 4 of (\ref{1b2}), the case $\beta = 4$, which can be written
\begin{equation}\label{NN1}
\Big ( {2 \over \pi} \Big )^N \prod_{l=1}^N (1 - | z_l|^2) \chi_{|z_l| < 1}
\prod_{1 \le j < k \le N} | z_k - z_j|^2 |1 - z_j \bar{z}_k |^2,
\end{equation}
has a plasma interpretation. We consider a one-component plasma formed by placing $N$ unit charges in the unit
disk with Neumann boundary conditions. We know (see e.g.~\cite[\S 15.9]{Fo10}) that in this setting  a charge at point
$z'$ effectively creates an image particle of identical charge at the point $1/\bar{z}'$, and thus the pair potential at point $z$
is
\begin{equation}\label{NN1a}
- \log ( |z - z'| | 1 - z z'|).
\end{equation}

Consider now a one component plasma system consisting of $N$ particles of unit charge interacting via the pair
potential (\ref{NN1a}), and coupled to a background charge density $-\eta$. Due to the particles being restricted to
a finite volume (the disk), the background charge density does not have to be neutralising for the particles to remain
confined. A short calculation, the result of which is reported in \cite[Eq.~(15.189)]{Fo10}, gives that the total energy of the system,
up to an additive constant, is equal to
$$
U = {1 \over 2} \sum_{j=1}^N \Big ( \pi \eta |z_j|^2 - \log (1 - |z_j|^2) -
\sum_{1 \le j < k \le N} \log \Big ( | z_k - z_j| |1 - z_j \bar{z}_k| \Big ).
$$
Thus we see that the eigenvalue PDF (\ref{NN1}) result as the Boltzmann factor $e^{-\beta U}$ of this system with $\beta = 2$,
provided we set $\eta = 0$, and thus there is no background charge.

In \cite[\S 15.9]{Fo10} the correlation functions for this same plasma system, except that $\eta$ was chosen so that the
background is neutralising, was computed explicitly. The working therein requires only minor modification to allow for
$\eta = 0$. The eigenvalues form a Pfaffian point process \cite[Ch.~6]{Fo10}. The one-point function is given by
 \begin{equation}\label{NN2}  
 \rho_{(1)}(z) = {1 \over \pi}\Big ( h(|z|^2) - |z|^{4N-2} h(|z|^{-2}) \Big ), \qquad h(s) = {d \over ds} {1 - s^{N+1} \over 1 - s}.
 \end{equation}
 In the limit $N \to \infty$ with no scaling of the eigenvalues, the Pfaffian reduces to the determinant
 \begin{equation}\label{NN2}
 \rho_{(k)}(z_1,\dots,z_k) = \pi^{-n} \det \Big [ {1 \over (1 - z_j \bar{z}_l)^2} \Big ]_{j,l=1,\dots,k}.
 \end{equation}
 This is well known \cite{PV03} as the correlation function, in the $N \to \infty$ limit, of the zeros inside the unit disk
 the random complex polynomials $\sum_{n=0}^N \alpha_n z^n$, where each $\alpha_n$ is an independent
 standard complex Gaussian, and has the property of being  invariant with respect to M\"obius transformations
 which map the unit disk to itself. The correlations (\ref{NN2}) are the same as those known for (\ref{1b2}) in
 the case $\beta = 2$ \cite{Kr06}, \cite[Ex.~15.7 q.2(iv)]{Fo10}, which corresponds to the one-component plasma system in a disk
 with no background charge.
 
 A further point of interest is to enquire if an explicit functional form analogous to (\ref{1b2}) for the ensemble obtained by truncating
 $n$ real quaternion rows and columns of $(N+n) \times (N+n)$ CSE, and so being left with an $N \times N$ real quaternion sub-block,
 can be obtained for general $n,N$? To answer this question, we note that CSE matrices can be characterised as self-dual  matrices
 further constrained to be unitary. Furthermore the fact that the unitary matrices are to be chosen with Haar measure
 tells us that the underlying matrices are Gaussian --- Haar measure results from orthogonlising a basis
 of Gaussian vectors. For $n$ large and $N$ fixed we thus expect that an $N \times N$ self-dual  sub-block can
 be well approximated to have Gaussian entries. Such an effect is well known for truncations of
 unitary or orthogonal matrices \cite{Ji06}, as can be seen by scaling $G \mapsto {1 \over \sqrt{n} }G$ in (\ref{2.40a}) with $n = N$,
 $M \mapsto n + N$, $\beta = 1$ or 2 and taking $n \to \infty$. But we know from the discussion of Section \ref{S4} that the
 ensemble of Gaussian self-dual matrices does not allows for explicit determination of its eigenvalue PDF, so
 we conclude that the sought generalisation of (\ref{1b2}) is not possible.
 
 \section*{Acknowledgements}
 This research was supported by the Australian Research Council grant DP140102613.
 I thank Jesper Ipsen for useful feedback relating on an earlier draft. Further, I thank
 Dong Wang and Jac Verbaarschot at NUS, and the Simons Center for Geometry and Physics
 (program on Random Matrices, fall semester 2015) respectively, for their hospitality and arranging financial support during my study leave period
 when ideas for the present paper were being formulated. 
 
%\bibliographystyle{amsplain}
%\bibliography{book1t}

\providecommand{\bysame}{\leavevmode\hbox to3em{\hrulefill}\thinspace}
\providecommand{\MR}{\relax\ifhmode\unskip\space\fi MR }
\providecommand{\MRhref}[2]{%
  \href{http://www.ams.org/mathscinet-getitem?mr=#1}{#2}
}
\providecommand{\href}[2]{#2}

\end{document}